\documentclass{article}

\usepackage{PRIMEarxiv}

\usepackage[utf8]{inputenc} 
\usepackage[T1]{fontenc}    
\usepackage{hyperref}       
\usepackage{url}            
\usepackage{booktabs}       
\usepackage{amsfonts}       
\usepackage{nicefrac}       
\usepackage{microtype}      
\usepackage{lipsum}
\usepackage{fancyhdr}       
\usepackage{graphicx}       
\graphicspath{{media/}}     
\usepackage{algorithm}
\usepackage{algpseudocode}
\usepackage{amsmath}
\usepackage{ragged2e}
\usepackage{subcaption}

\title{A self-supervised learning framework for seismic low-frequency extrapolation
}

\author{
  Shijun Cheng, Randy Harsuko, Tariq Alkhalifah \\
  King Abdullah University of Science and Technology, Thuwal 23955-6900, Saudi Arabia. \\
  \texttt{sjcheng.academic@gmail.com, \{mochammad.randycaesario, tariq.alkhalifah\}@kaust.edu.sa} \\
   \And
  Yi Wang, Qingchen Zhang \\
  Innovation Academy for Precision Measurement Science and Technology, \\ Chinese Academy of Sciences, Wuhan 430077, China. \\
   \texttt{\{wangyi, qczh\}@apm.ac.cn} \\
}

\begin{document}
\maketitle

\begin{abstract}
Full waveform inversion (FWI) is capable of generating high-resolution subsurface parameter models, but it is susceptible to cycle-skipping when the data lack low-frequency. Unfortunately, the low-frequency components (\(<\) 5.0 Hz) are often tainted by noise in real seismic exploration, which hinders the application of FWI. To address this issue, we develop a novel self-supervised low-frequency extrapolation method that does not require labeled data, enabling neural networks to be trained directly on real data. This paradigm effectively addresses the significant generalization gap often encountered by supervised learning techniques, which are typically trained on synthetic data. We validate the effectiveness of our method on both synthetic and field data. The results demonstrate that our method effectively extrapolates low-frequency components, aiding in circumventing the challenges of cycle-skipping in FWI. Meanwhile, by integrating a self-supervised denoiser, our method effectively performs simultaneously denoising and low-frequency extrapolation on noisy data. Furthermore, we showcase the potential application of our method in extending the ultra-low frequency components of the large-scale collected earthquake seismogram.

\end{abstract}

\section*{Plain Language Summary}

Full waveform inversion (FWI) is a method used to provide detailed underground images for efforts in oil exploration or studying earthquakes. However, the method is prone to a problem known as "cycle-skipping", caused by the lack of low frequencies in the data, and as a result, the inversion converges to an inaccurate velocity model. We propose a new way to enhance the low-frequency components using a neural network-based the self-supervised approach. This means our method learns directly from real data, rather than relying on artificially created data, which is a common limitation in the supervised paradigm. Our method not only helps to overcome the issue of skipping cycles but also demonstrates robustness against noisy data, enhancing its practical application potential. The tests on exploration data validate its ability to predict the low-frequency components, helping to avoid local minima and thus improving the accuracy of FWI. Also, we demonstrate how our method can be applied to invert earthquake data, where it can extend the ultra-low frequency information. This research could contribute to providing a better and a more reliable way to obtain images of the Earth's interior.

\section{Introduction}
 Full waveform inversion (FWI) can provide high-resolution subsurface parameter model by minimizing the misfit between observed and modeled seismic waveforms. This approach has drawn a lot of attention \cite{ben2011efficient,choi2012application,anderson2012time,zhang2014efficient,zhang2016robust,guo2017elastic,zhang2018hybrid} since it was proposed by Tarantola \cite{tarantola1984inversion,tarantola1986strategy}. However, one common problem for FWI is the high nonlinearity of the problem, resulting in many local minima, which can prevent the algorithm from converging to the global minimum. To address this problem, various strategies and algorithms have been developed to help eliminate or escape local minima in FWI. 
	
Bunks et al. \cite{bunks1995multiscale} introduced an multi-scale strategy that initiates with lowest frequencies, and then progressively adds higher frequencies to the inversion. This method, applied widely, attempts to help FWI avoid local minima. It accomplishes this goal by smoothing the objective function in the early stages and subsequently allows higher resolution information as the inversion advances. Following this work, Boonyasiriwat et al. \cite{boonyasiriwat2009efficient} provided an optimal multi-scale frequency selection algorithm to achieve computational efficiency in the time domain. Besides, Gao et al. \cite{8444461} and Ren and Liu \cite{ren2015elastic} also attempted to refine the multi-scale algorithm to improve the inversion quality of FWI. Nevertheless, these methods do not address the issue of the lack of low frequencies in the data.
	
In order to build the long-wavelength components of the subsurface parameter model in the case of lacking low-frequencies,  Ha and Shin \cite{ha2012laplace} and  Kim et al. \cite{kim2013algorithm} proposed the Laplace-domain waveform inversion. However, this method has the drawback that the penetration depth of the Laplace-domain inversion. the  depends on the offset and the choice of Laplace damping constants. Hence, numerous studies aimed to adjust the norm of the misfit function to minimize the occurrence of local minima. Ma and Hale \cite{ma2013wave} utilized the dynamic warping algorithm, mitigating FWI's reliance on an accurate initial model by estimating the travel time shift between observed and synthetic data. Pointed out by Wu et al. \cite{wu2014seismic} and Luo and Wu \cite{luo2015seismic}, the seismic data's envelope potentially holds ultra low-frequency signals, offering an avenue to estimate long-wavelength velocity structures even in the absence of low-frequency information. Furthermore, Warner and Guasch \cite{warner2016adaptive} introduced the adaptive waveform inversion (AWI) to counter the cycle-skipping problem. This method employs a convolutional filter, transforming predicted data into observed data. Yang and Engquist \cite{yang2018analysis} and Yang et al. \cite{yang2018application} explored the optimal transport approach, measuring amplitude differences and global phase shifts, offering a solution to circumvent cycle-skipping issues. Sun and Alkhalifah \cite{sun2019application} showed that AWI is part of a more general framework based on matching filters and optimal transport. Leveraging well logging data, Li et al. \cite{li2022well} devised a structure-guided interpolation algorithm to construct the background velocity model for FWI. Yao and Wang \cite{yao2022building} crafted a full-waveform inversion starting model from wells, employing dynamic time warping and convolutional neural networks. Another prominent technique for velocity model construction is the reflection waveform inversion (RWI) \cite{xu2012full}, requiring migration and demigration to update the wavepath. RWI typically incorporates a correlation-type objective function to address amplitude challenges encountered without true amplitude \cite{chi2015correlation,alkhalifah2016multiscattering,wu2017efficient,guo2017elastic,wang2018elastic,zhang2019local,yao2020review}. However, RWI demands significantly higher computational resources compared to conventional FWI methods. 

Recently, machine learning has been demonstrated as significant potential in the field of seismic processing due to its ability to approximate any nonlinear function \cite{wu2019faultseg3d, yu2019deep, mousavi2022deep, harsuko2022storseismic, cheng2023elastic, cheng2023meta}, which also showed a potential in velocity model building. As a result, some researchers have proposed replacing traditional physics-based FWI with convolutional neural networks (CNNs) to directly learn a mapping from seismic data to velocity models in a data-driven manner \cite{araya2018deep, 8931232, du2022deep}. These trained CNN models can then be applied directly to unseen test data to predict velocity models. Typically, the training of such CNNs requires a large dataset and is conducted in a supervised learning (SL) manner. For example, Yang and Ma \cite{yang2019deep} employed the classic U-Net network architecture as a baseline, trained it on a self-constructed dataset, and then tested it on 2D simulated models and the SEG salt model, demonstrating its potential in constructing velocity models. Wu and Lin \cite{wu2019inversionnet} integrated a CNN with a conditional random field to form a hybrid network model, aiming to enhance the network's ability to refine the velocity fields near faults and boundaries. Zhang and Gao \cite{zhang2021deep} suggested the use of images in the common-shot domain, replacing shot gathers, and then established an iterative deep CNN to produce high-resolution velocity models. Yang et al. \cite{yang2023well} also utilized low-resolution velocity models, images, and well velocity constraints to reconstruct high-resolution velocity models. While these data-driven approaches offer a new direction for seismic inversion, they often encounter challenges in generalization due to the complexity of real subsurface structures and property distributions. Moreover, the constructed training datasets often fail to cover real underground features. To address the lack of physical interpretability of purely data-driven approaches, some researchers have proposed integrating a forward simulator based on the wave equation into the neural network (NN)-based seismic inversion \cite{jin2021unsupervised, ren2021seismic, lin2023physics}. This simulator is employed to forward model the subsurface properties predicted by the NN. The discrepancy between the resulting seismic records and the original observations is then used as a misfit to update the network. This approach not only eliminates the need for labeled data but also enhances generalization to some extent. However, these methods have still not achieved satisfactory results on field data and often face challenges, such as unclear inversion boundaries and overly blurred background velocities in situations when low-frequencies are absent \cite{yang2019deep}. 

Instead of using NNs to approximate an inversion operator, can we leverage their low-frequency extrapolation capabilities, and thus, extend the low-frequency information of observed data. Compared to the daunting task of directly approximating the inversion operator with NNs, using them to learn low-frequency extension might be more realistic. Some researchers have initiated preliminary explorations and applied this concept within FWI. Ovcharenko et al. \cite{ovcharenko2017neural} were among the first to suggest using a feed-forward artificial neural network (ANN) to approximate the nonlinear relationship between frequency-wavenumber spectra with limited bandwidth and the low-frequency data and, thus, recovering low-frequency information. To tackle the inherent challenge of ANNs, wherein the number of trainable parameters surges with increasing input data, Ovcharenko et al. \cite{ovcharenko2019deep} further transitioned from ANNs to CNNs and applied it for low-frequency extrapolation in multi-offset seismic data. Instead of working in the frequency-wavenumber domain, Sun and Demanet \cite{sun2018low, sun2020extrapolated} proposed using an CNN to directly reconstruct low-frequency components trace-by-trace from band-limited data in the time domain. Furthermore, Sun and Demanet \cite{sun2021deep} extended their proposed framework to the low-frequency extrapolation of multicomponent elastic wave data. Hu et al. \cite{hu2021progressive} and Jin et al. \cite{jin2021efficient} developed progressive transfer learning algorithms, which embed a physics-based FWI module to iteratively update the training dataset, thereby reducing the feature discrepancy between it and the test dataset. These studies have provided validation of the feasibility of using NNs for seismic data's low-frequency extrapolation. However, they only tested on synthetic data and rarely showcased applications on field data. Thus, the performance on field data is our ultimate goal.

Achieving good low-frequency extension results on field data using data-driven methods remains a challenge. Consequently, relatively little attention has been devoted to enhancing the performance of NNs in low-frequency extrapolation on field data. Fabien-Ouellet \cite{fabien2020low} utilized recurrent CNN to simultaneously achieve seismic denoising and low-frequency generation, validating the method's performance on both synthetic and field data. Additionally, Nakayama and Blacquiere \cite{nakayama2021machine} developed a multi-task seismic processing framework based on NNs, capable of concurrently suppressing the blending noise, interpolating missing traces, and recovering low-frequency information. Fang et al. \cite{fang2020data} employed a CNN architecture based on convolutional autoencoders to learn the relationship between low-frequency and high-frequency data, thereby predicting low-frequency components from high-frequency data. They showcased the effectiveness of low-frequency recovery in land field data and conducted a comparative analysis between FWI and RTM products using both the original dataset and the dataset enhanced with NN-predicted low-frequency components. In a similar vein, Wang et al. \cite{wang2022low} applied a dense CNN methodology to broaden the low-frequency spectrum of prestack viscoacoustic seismic data, yielding promising outcomes when tested on marine field data. Ovcharenko et al. \cite{ovcharenko2022multi} developed a multi-task processing framework, which not only recovers low-frequency information but also provides a smooth subsurface background model. This was applied to marine data involving the elastic assumption. Although these methods have demonstrated a certain ability in recovering low frequencies on field data, their performance significantly lags behind their results on synthetic data. This discrepancy is attributed to their reliance on an SL approach where training occurs solely on synthetic data before being directly applied to field data. It is evident that field and synthetic data have considerable feature differences due to physical approximations made during synthetic data generation. Consequently, the features captured by the NN on synthetic data do not generalize well when applied to field data. To bridge the feature gap between the synthetic and field data, Alkhalifah et al. \cite{alkhalifah2022mlreal} introduced a domain-adaptive method, namely MLReal, which embeds real features of field data into synthetic data to enhance low-frequency extrapolation performance on field data. However, this preprocessing might eliminate some characteristics of seismic data during the conversion, such as phase. Therefore, a more viable alternative is to train directly on field data. It allows the network to extract frequency features from field data more directly, and thus, contributes to low-frequency tasks on field data.

Indeed, predicting low frequency data using unsupervised learning poses significant challenges. To our knowledge, we have only come across two studies addressing this issue. One study by Sun et al. \cite{sun2023learning} developed a semi-supervised learning approach by incoorperating a domain adaptation approach. They trained a generative model, CycleGAN, using synthetic low-frequency shot gathers and paired field band-limited shot gathers. The trained CycleGAN demonstrated its capability to extrapolate low-frequency components during testing on field data. However, during its training phase, CycleGAN still captures features inherent to the synthetic data, which might pose generalization issues when predicting field data. For instance, field data may exhibit anisotropic effects, but synthetic data simulated under the isotropic assumption may not accurately represent the field data. Meanwhile, training CycleGANs is not trivial. Since it requires simultaneous training of both discriminator and generator models, it relies on expert tuning and hyperparameter selection. Another work was proposed by Wang et al. \cite{wang2020self}, employing a self-supervised learning (SSL) method for low-frequency extension in seismic data. They first trained an NN using paired seismic data sets missing the low-frequency components. These paired data sets share the same high-frequency cutoff but differ in their low-frequency components. The dataset with the relative low-frequency component, sourced from original data, serves as labels. In contrast, the data devoid of the relative low-frequency components, achieved by filtering out the low-frequency part from the original data, acts as the NN's input. Subsequently, they employed the trained network to generate low-frequency components. However, since the training data for this method only extracts labels from the original data missing low frequencies, its performance in low-frequency reconstruction relies on the low-frequency range of the original seismic data. Moreover, downsampling significantly reduces the time sampling length of the seismic data, leading to a loss of high-frequency information and a consequent decline in time resolution.

In line with the work by Wang et al. \cite{wang2020self}, our study introduces a Lesslow2Low (L2L) framework. This approach involves applying a high-pass filter to the initial observed seismic data, which already lacks low-frequency components, thereby creating input data with even fewer low-frequency contents. The original observed data serve as pseudo-labels. In fact, the initial inspiration for L2L came from the classic denoising algorithm Noisier2Noise \cite{moran2020noisier2noise}, which adds noise to the original noisy data to create data with stronger noise. This noisier data is considered as input, while the original noisy data acts as pseudo-labels, allowing the network to be trained in an SSL fashion. The subtle distinction between the L2L framework and the method of \cite{wang2020self} lies in our ability to apply high-pass filters with varying cutoff frequencies during the epoch progression, as opposed to a fixed approach. However, the L2L framework also faces similar problems: the network's low-frequency extrapolation performance is limited by the low-frequency range of the original observed data. We present an example that substantiates this point and leads to two important findings: 1. The reason for the L2L framework's limited performance, compared to the SL method, is the frequency information bias between the training set employed in L2L and that used in the SL method. 2. The NN, which is trained using the L2L framework, demonstrates a certain degree of low-frequency extension capability with the original observed data. Although this capability is limited, if we can iteratively refine the frequency information bias between the original observed data and the ideal ground truth, we can gradually approach the low-frequency extrapolation performance of the SL method.

Motivated by these findings, we develop an effective SSL low-frequency extrapolation algorithm and apply it to FWI. In our approach, the NN training is divided into two stages: warm-up and iterative data refinement (IDR). The difference between these stages lies in the training sets used. During the warm-up stage, the original observed data acts as the pseudo-labels, and the input data is generated by applying a high-pass filter to the observed data. In the IDR stage, the pseudo-labels for the current epoch are derived from the predictions made by the network trained in the previous epoch on the original observed data, and the input data are obtained by high-pass filtering these predicted pseudo-labels. The warm-up stage, which precedes the IDR stage, aids in stabilizing the network and initially capturing data characteristics, thus providing some degree of low-frequency extrapolation capability. The IDR stage gradually narrows the gap between the predicted pseudo-label and the ideal ground truth, thereby enhancing the network's low-frequency extrapolation performance. We first conduct tests on synthetic data to validate the effectiveness of our method and the ability to avoid cycle-skipping local minima. Subsequently, we explore the application potential of our method in exploration field data to improve the accuracy of FWI. Lastly, we implement our algorithm on extensive earthquake seismogram data to conduct low-frequency extrapolation testing, assessing its capability to extend the spectrum to extremely low-frequency components.

\section{Method}
\subsection{FWI}
Using the whole information of seismic data, FWI can construct a high-resolution subsurface model by iteratively minimizing the misfit between observed and simulated data. The conventional misfit objective function is given by the $l_2$ norm as follows:
\begin{equation}\label{eq1}
\setlength{\abovedisplayskip}{5pt}
\setlength{\belowdisplayskip}{5pt}
		\chi(\textbf{m})=\frac{1}{2} \left\| d^{obs}(\textbf{x}_r,\textbf{x}_s)-d^{syn}(\textbf{x}_r,\textbf{x}_s) \right\| ^2_2,
\end{equation}
where the superscripts $obs$ and $syn$ denote the observed and simulated seismic data, respectively, $\textbf{x}_r$ and $\textbf{x}_s$ represent the coordinates of the receivers and sources, respectively.
FWI attempts to find a medium capable of generating simulated data that closely align with the observed ones. Since FWI is a waveform matching process, there would be mismatching problems between the simulated and observed seismograms when the initial model is poor, leading to a time delay greater than $T/2$ (where $T$ is the period of a wavelength), that is, the so-called cycle-skipping problem. The poor initial model often leads FWI to local minima.

In order to overcome this problem, researchers have proposed many of the aforementioned algorithms, mostly by modifying the misfit function norm \cite{ma2013wave,wu2014seismic,warner2016adaptive,yang2018analysis,yong2023localized}. Actually, the root cause of the cycle-skipping problem is the velocity model lacking the long-wavelength components, which is often constructed with the low-frequency data information \cite{virieux2009overview}. Unfortunately, in real seismic exploration, data with frequencies below 5 Hz are usually contaminated by noise, thus rendering unusable low-frequency data. Unlike the aforementioned methods, we try to predict the low-frequency data information for the observed data using an innovative machine learning algorithm. 

To enhance our focus on the phase information and to avoid the amplitude challenge, we adopt a convolution-based misfit function \cite{choi2011source}, which has a similar performance to the correlative norm in RWI.
\begin{equation}\label{eq2}
\setlength{\abovedisplayskip}{5pt}
\setlength{\belowdisplayskip}{5pt}
		\chi(\textbf{m})=\frac{1}{2} \left\| d^{obs}(\textbf{x}_r,\textbf{x}_s)\ast d^{syn}(x_{ref},\textbf{x}_s)-d^{syn}(\textbf{x}_r,\textbf{x}_s)\ast d^{obs}(x_{ref},\textbf{x}_s) \right\| ^2_2,
\end{equation}
where the subscript $ref$ denotes a reference trace, which is selected from the near-offset traces with a high signal-to-noise ratio. Moreover, to improve the robustness against background noise, we further introduce a hybrid-norm objective function into equation \eqref{eq2}.
\begin{equation}\label{eq3}
\setlength{\abovedisplayskip}{5pt}
\setlength{\belowdisplayskip}{5pt}
		\chi_{hybrid} \left( \textbf{m} \right)=\sqrt{1+\frac{\left\| \triangle \hat{d}(\textbf{x}_r,\textbf{x}_s) \right\|^2 }{\epsilon^2}}-1,
\end{equation}
where $\triangle \hat{d}(\textbf{x}_r,\textbf{x}_s)=\hat{d}^{obs}(\textbf{x}_r,\textbf{x}_s)-\hat{d}^{syn}(\textbf{x}_r,\textbf{x}_s)$, $\hat{d}^{obs}$ and $\hat{d}^{syn}$ represent the first and second convolution terms in equation \eqref{eq2}. As for the damping coefficient, we follow the criterion as
\begin{equation}\label{eq4}
\setlength{\abovedisplayskip}{5pt}
\setlength{\belowdisplayskip}{5pt}
		\epsilon=c\cdot mean(\left|\hat{d}^{obs}\right|),
\end{equation}
where $c$ is a constant ranging between $0.1\sim100.0$. The smaller $c$ is the greater the added gain to the weak signal, which helps enhance the robustness of FWI to noise.

\subsection{Supervised low-frequency extrapolation}
When seismic waves travel through the Earth's interior, the stratigraphic filtering effect restricts the frequency content of the observed data. This restriction is especially obvious at lower frequencies, resulting in a conspicuous absence of low-frequency components in the observed data. Therefore, the recorded seismic data can be expressed as
\begin{equation}\label{eq5}
\setlength{\abovedisplayskip}{5pt}
\setlength{\belowdisplayskip}{5pt}
d^{obs} = H[d^{real}],
\end{equation}
where $d^{obs}$ represents the observed data with an absence of low-frequency components, $H[\cdot]$ symbolizes the high-pass filter effect of the Earth's strata on the seismic waves, and 
$d^{real}$ denotes the broad band ground-truth data. 

The feasibility of an NN-based data-driven approach for low-frequency extrapolation has been validated \cite{ovcharenko2019deep, sun2020extrapolated}. Generally, we can employ the SL technique to train an NN that provides a non-linear mapping. This mapping translates data restricted in frequency bandwidth, particularly lacking in low-frequency components, to a dataset that encompasses the low-frequency components. We can represent this operation as follows: 
\begin{equation}\label{eq6}
\setlength{\abovedisplayskip}{5pt}
\setlength{\belowdisplayskip}{5pt}
d^{real} = \text{NN}(d^{obs}, \boldsymbol{\theta}),
\end{equation}
where $\boldsymbol{\theta}$ is the learnable parameters of network $\text{NN}$. Due to the unavailability of labels in field data, a synthetic dataset is typically constructed to facilitate the training of the NN.

\subsection{LessLow2Low}
Compared to training on synthetic data, direct training on real data can significantly enhance the generalizability of NNs, thereby yielding superior low-frequency extrapolation on field data. However, real data typically lacks low-frequency information, which raises the question: how can we conduct effective training under these constraints? Inspired by a denoising method from the machine learning community, known as Noisier2Noise \cite{moran2020noisier2noise}, we can devise a strategy, called LessLow2Low (L2L), that facilitates training directly on real data. Within the Noisier2Noise framework, we are equipped solely with the original noisy data, which serves as a pseudo-label. The input data is generated by introducing additional noise to the already noisy original data. By constructing a noisier-noisy dataset in this manner, we can provide a substrate for training the NN. Analogously, we, in our L2L framework, assume we only have a waveform dataset that lacks low-frequency components, in which such data can be treated as pseudo-label data. Subsequently, a high-pass filter is applied to this pseudo-label data to obtain the input for the NN. In this context, the high-pass filtered data, relative to the original waveform data, possess less low-frequency components, thereby enabling the establishment of an SSL regime on real data. Nevertheless, considering the higher nonlinearity in the relation between low and high frequency data, this paradigm often provides limited capabilities for low-frequency extrapolation. In the following, we will present an example to illustrate this point.

We simulate synthetic data for the Marmousi2 model with a dominant frequency of 15 Hz, followed by the preparation of three distinct training datasets. These datasets share the same input data, derived by subjecting the simulated synthetic data to a high-pass filter with a cutoff of 10 Hz. However, the label data for each set are unique. For the first dataset, the label data come from the original synthetic dataset, designated as low-full. The second dataset's label data are obtained by applying a high-pass filter with a cutoff frequency at 5 Hz to the original synthetic data, identified as lesslow-low 1. The third dataset's label data are filtered to remove frequencies below 7 Hz from the original synthetic data, termed lesslow-low 2. The first dataset serves as the foundation for the SL of the NN. The rationale behind constructing the latter two datasets is to consider the reality that real data often lacking low-frequency components. We train NNs using these three datasets, all under the same training configuration as elucidated in subsequent sections.

\begin{figure*}[!t]
\centering
\includegraphics[width=0.4\textwidth]{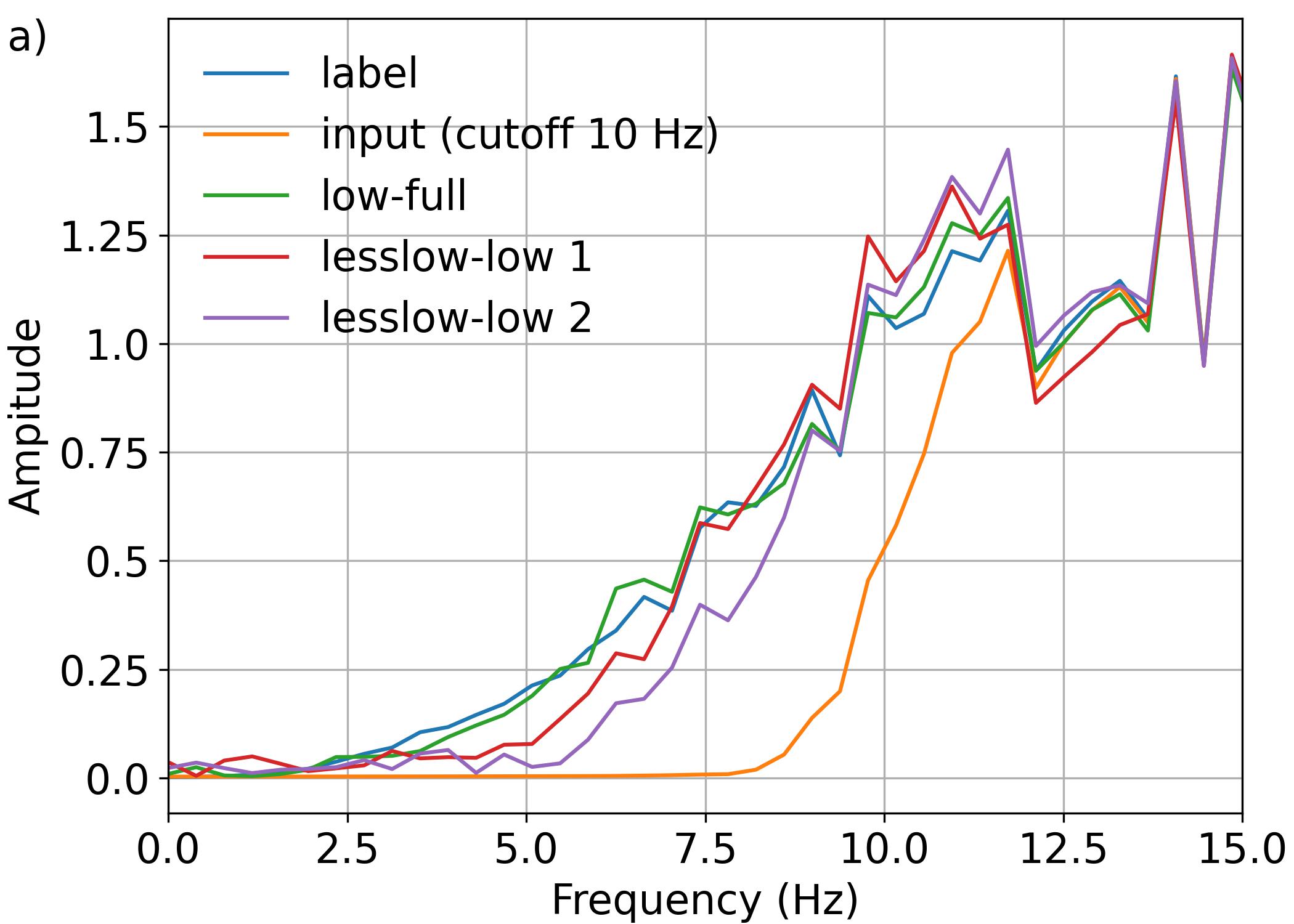}
\includegraphics[width=0.4\textwidth]{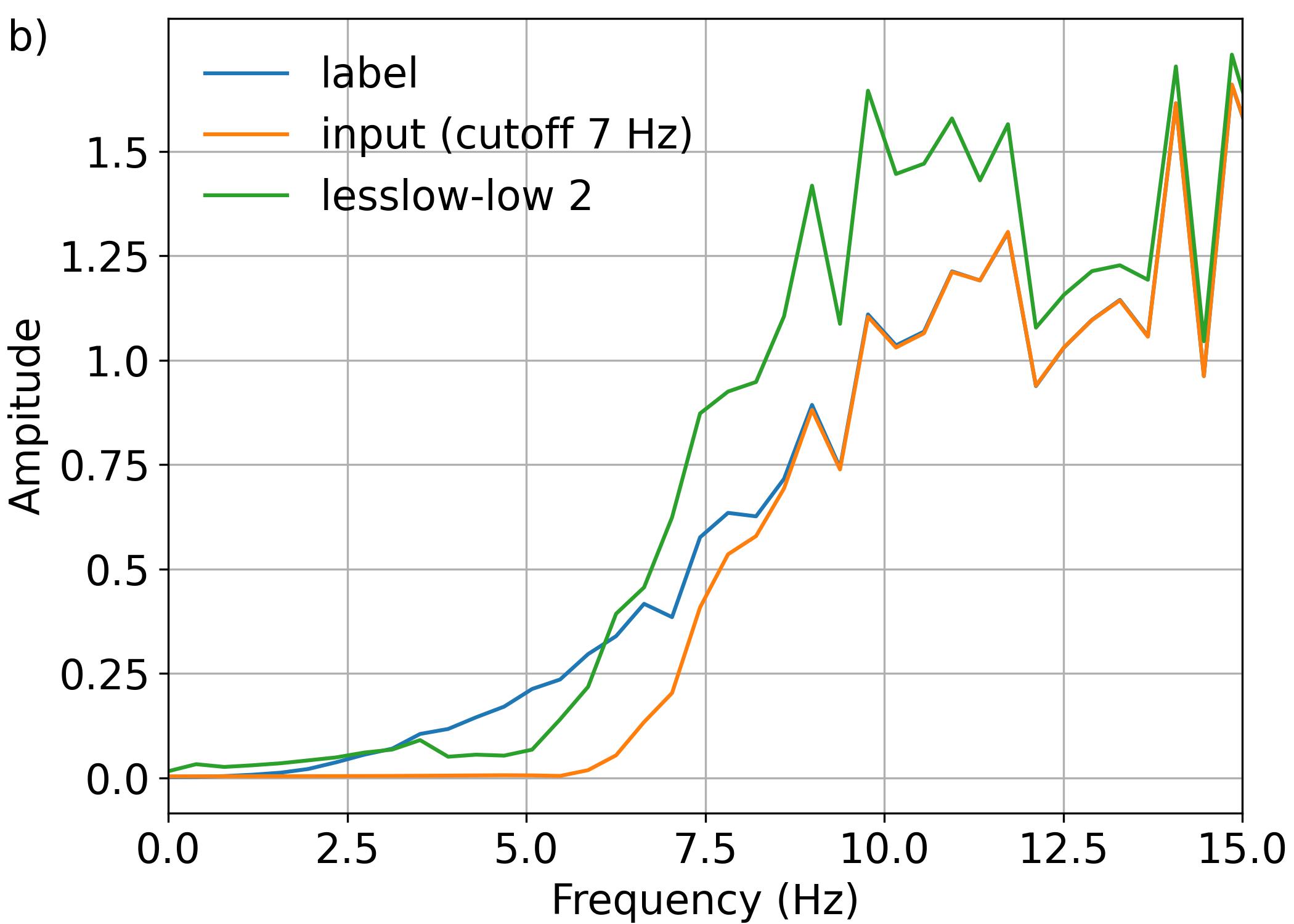}
\caption{(a) Comparison of amplitude spectrum curves from the prediction results of networks trained on different datasets. (b) The prediction results of the network trained on the lesslow-low 2 dataset for the original input data, which lack frequency components below 7 Hz, are compared with the amplitude spectrum curves of the input data and their corresponding labels.}
\label{fig1}
\end{figure*}

We employ a test dataset, from which frequencies below 10 Hz have been filtered out, to assess the low-frequency extrapolation capabilities of networks trained on three distinct training sets. We extract a single trace from their prediction results for the test data and plot the corresponding amplitude spectra (see Figure \ref{fig1}a), comparing them with the test data and their associated labels. We observe that the network trained on the low-full dataset achieves superior low-frequency extrapolation, which is definitively attributed to the network being trained in an SL fashion. In contrast, the NNs trained on the two lesslow-low datasets exhibit somewhat diminished performance. This substantiates the preceding contention that a training paradigm which uses data lacking low-frequency components as labels can only furnish a limited capacity for low-frequency extrapolation. For example, the label data in the lesslow-low 1 dataset are devoid of frequency components below 5 Hz. Therefore, they do not include information in their prediction distribution for frequencies below 5 Hz contained the test data. An additional finding is that the network, trained on lesslow-low 1, exhibits better low-frequency extrapolation performance than the network trained on lesslow-low 2. This can be attributed to the fact that the lesslow-low 1 dataset has a lower frequency information bias relative to the lesslow-low 2 dataset when compared with the low-full dataset. 

An additional test involving using the network trained on the lesslow-low 2 dataset to predict the corresponding pseudo-label data, which lack below 7 Hz frequencies. The single-trace amplitude spectrum from the prediction is compared with the pseudo-label data, as well as with the original synthetic data, which is depicted in Figure \ref{fig1}b. We can see that the network trained on lesslow-low data is capable of extending the frequency to a certain extent from the original pseudo-label data.

The two illustrative examples impart two key insights: First, if we can mitigate the frequency information bias between the lesslow-low and low-full datasets, we can incrementally approach the frequency extrapolation performance of a network trained on the low-full dataset. Second, the networks trained on lesslow-low datasets exhibit a certain capacity to extrapolate lower frequencies in the pseudo-label data involved in their training. From these insights, we can infer that if we iteratively refine the low-frequency component information of the pseudo-label data using the network trained on the lesslow-low data, we could ultimately approximate the frequency extrapolation performance of an SSL framework on real data. Motivated by this inference, in the following section, we will present our SSL framework for low-frequency extrapolation.

\subsection{Self-supervised low-frequency extrapolation}
Drawing on the key findings from the previous section, we develop an SSL framework for low-frequency extrapolation. Our framework principally consists of two components: a warm-up and an iterative data refinement (IDR) phases. The detailed algorithmic procedure is delineated in Algorithm 1. In the following, we will elucidate the key components therein.

Firstly, the NN undergoes a warm-up phase. We begin by simply employing an L2L procedure to apply a high-pass filter to the original seismic data $\{x_i\}_{i=1}^N$, which lacks low-frequency components, thereby creating a lesslow-low dataset $\{H[x_i],x_i\}_{i=1}^N$. Subsequently, the NN is subjected to multiple epochs of optimization training on this dataset. The objective of this process is to enable rapid stabilization of the neural network, allowing it to preliminarily capture the characteristics of seismic data. Furthermore, the second finding from the previous section indicates that this pre-trained network, denoted as $\text{NN}_w$, exhibits a certain degree of low-frequency extrapolation ability with respect to the original label data. This capability forms the groundwork for the iterative refinement of the training set in subsequent stages.

Subsequently, the NN enters the IDR phase. In this stage, we first leverage the pre-trained network $\text{NN}_w$ to predict the original seismic data. The predictions serve as the initial pseudo-labels for the IDR phase, with corresponding inputs derived from applying a high-pass filter to these predictions. This procedure facilitates the creation of a new lesslow-low dataset, employed for the first epoch of training during the IDR stage:
\begin{equation}\label{eq7}
\setlength{\abovedisplayskip}{5pt}
\setlength{\belowdisplayskip}{5pt}
\text{NN}_0 \leftarrow \{(H[\text{NN}_w(x_i)], ~ \text{NN}_w(x_i))\}_{i=1}^N,
\end{equation}
where the network $\text{NN}_0$ is directly initialized from the per-trained network $\text{NN}_w$.

Informed by the first insight from the previous section, we anticipate that after training for one epoch on the new training set, the low-frequency extrapolation capability of $\text{NN}_0$ will slightly surpass that of $\text{NN}_w$. This improvement is attributed to the fact that, compared to the lesslow-low dataset used during the warm-up phase, the new lesslow-low dataset exhibits lower frequency representation in the labels compared to the low-full dataset. During the subsequent training, we iteratively perform this process to progressively diminish the frequency information bias between the lesslow-low and low-full datasets. Specifically, for each training epoch, we first employ the network trained in the previous epoch (e.g., $\text{NN}_{j-1}$) to predict the original data $\{x_i\}_{i=1}^N$, and thus, obtain the frequency-extended pseudo-labels $\{\text{NN}_{j-1}(x_i)\}_{i=1}^N$. Then, we apply a high-pass filter to these pseudo-labels to generate corresponding input data $\{H[\text{NN}_{j-1}(x_i)]\}_{i=1}^N$. The resulting lesslow-low training set $\{(H[\text{NN}_{j-1}(x_i)], ~ \text{NN}_{j-1}(x_i))\}_{i=1}^N$ will optimize the network $\text{NN}_j$ for one epoch, for example,
\begin{equation}\label{eq8}
\setlength{\abovedisplayskip}{5pt}
\setlength{\belowdisplayskip}{5pt}
\text{NN}_j \leftarrow \{(H[\text{NN}_{j-1}(x_i)], ~ \text{NN}_{j-1}(x_i))\}_{i=1}^N.
\end{equation}
After conducting multiple epochs of training in the IDR phase, our network incrementally aligns with the low-frequency extrapolation capabilities of a model trained on the low-full dataset.

\begin{algorithm}
\caption{Self-supervised low-frequency extrapolation}\label{alg:Framwork}
\textbf{Input:} Raw band-limited seismic data $\{x_i\}_{i=1}^N$. \\
\textbf{Input:} Neural network model NN. \\
\textbf{Input:} High-pass filter $H[\cdot]$. \\
\textbf{Input:} ${E_{warmup}}$: The number of epochs during the warm-up phase. \\
\textbf{Input:} ${E_{idr}}$: The number of epochs during the iterative data refinement phase. \\
\textbf{--------------------------------- Warm-up phase ------------------------------} \\
\textbf{Output:} Pre-trained model $\text{NN}_w$
\begin{algorithmic}
\State 1: Randomly initialize network parameters ${\boldsymbol{\theta}}$
\State 2: \textbf{for} ${j}$ \textbf{in} ${E_{warmup}}$ \textbf{do}
\State 3: \quad \quad Perform the high-pass filter on raw seismic data $H[x_i]$
\State 4: \quad \quad Construct the lesslow-low dataset $\{H[x_i],x_i\}_{i=1}^N$
\State 5: \quad \quad Optimize the network $\text{NN}_w$ on the lesslow-low dataset \\ \quad \quad \quad \quad \quad \quad \quad \quad \quad \quad \quad
$\text{NN}_w \leftarrow \{H[x_i],x_i\}_{i=1}^N$
\State 6: \textbf{end for}
\end{algorithmic}
\textbf{----------------------- Iterative data refinement phase ---------------------} \\
\textbf{Output:} Final low-frequency extrapolation model $\text{NN}_{E_{idr}}$
\begin{algorithmic}
\State 7: Predict the raw band-limited seismic data $\{\text{NN}_w(x_i)\}_{i=1}^N$
\State 8: Construct the new refined lesslow-low dataset \\ \quad \quad \quad \quad \quad \quad \quad \quad \quad \quad \quad \quad  $\{H[\text{NN}_w(x_i)],\text{NN}_w(x_i)\}_{i=1}^N$
\State 9: Initialize the network $\text{NN}_0 = \text{NN}_w$
\State 10: Optimize the network $\text{NN}_0$ on the new lesslow-low dataset for one epoch \\ \quad \quad \quad \quad \quad \quad \quad \quad \quad \quad \quad
$\text{NN}_0 \leftarrow \{H[\text{NN}_w(x_i)],\text{NN}_w(x_i)\}_{i=1}^N$ 
\State 11: \textbf{for} ${j \rightarrow 1}$ \textbf{in} ${E_{idr}}$ \textbf{do}
\State 12: \quad \quad Generate the low-frequency extended data $\{\text{NN}_{j-1}(x_i)\}_{i=1}^N$
\State 13: \quad \quad Construct the new refined lesslow-low dataset  \\ \quad \quad \quad \quad \quad \quad \quad \quad \quad \quad \quad \quad  $\{H[\text{NN}_{j-1}(x_i)],\text{NN}_{j-1}(x_i)\}_{i=1}^N$
\State 14: \quad \quad Initialize the network $\text{NN}_j = \text{NN}_{j-1}$
\State 15: \quad \quad Optimize the network $\text{NN}_j$ on the new lesslow-low dataset for one epoch \\ \quad \quad \quad \quad \quad \quad \quad \quad \quad \quad \quad
$\text{NN}_j \leftarrow \{H[\text{NN}_{j-1}(x_i)],\text{NN}_{j-1}(x_i)\}_{i=1}^N$ 
\State 16: \textbf{end for}
\end{algorithmic}
\end{algorithm}

\subsection{Network architecture and training procedure}
\begin{figure*}[!t]
\centering
\includegraphics[width=0.98\textwidth]{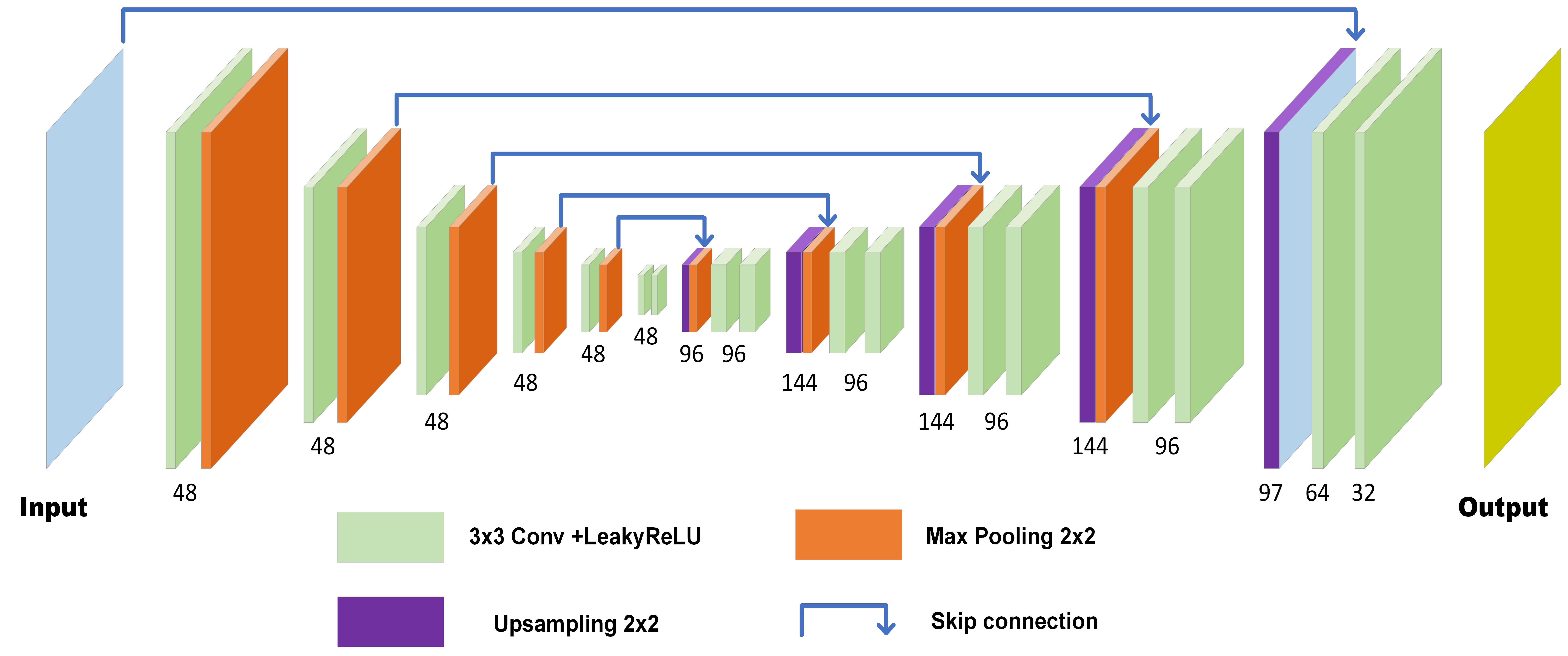}
\caption{The neural network architecture used in our study. }
\label{fig2}
\end{figure*}

Our SSL framework employs a classic network architecture, namely U-Net \cite{ronneberger2015u},  which has been widely utilized in the NN-based seismic processing workflows \cite{wu2019faultseg3d, yang2019deep, cheng2023meta}. Figure \ref{fig2} comprehensively details the network architecture utilized in this study. It's noteworthy that we did not strictly follow the classic U-Net structure, but instead make several modifications to suit our specific requirements. The first significant change is in scaling: rather than adopting the four scales typical of the traditional U-Net, we incorporate five scales, involving five 2x2 downsampling and 2x2 upsampling operations. The rationale behind this modification is that different scales often represent different frequency component information. By extracting features across multiple scales, we aim to train an NN that is more attuned to various frequency components, which in turn is expected to enhance the network's capability in low-frequency extrapolation. The second modification involves the skip connections at the network's maximum scale: they directly receive the input data, rather than data that have undergone transformations through an input layer. This strategy is adopted to circumvent the loss of original signal information that can be caused by the network's depth. Since CNNs could potentially create a smoothing effect, which might entail the loss of high-frequency signals, we deliberately avoid impairing these high-frequency signals during the process of low-frequency extrapolation. The third modification exists in the convolution layers: whereas the classic U-Net baseline employs a 3x3 convolution layer followed by batch normalization (BN) and a Leaky Rectified Linear Unit (LeakyReLU) activation function, our experiments revealed that incorporating BN would produce unstable low-frequency signals during the IDR phase. Therefore, we omit BN to ensure that the network provides a reliable solution. 

During the training process, we present a hybrid loss function to co-optimize the network. This hybrid loss function consists of a data loss $\mathcal{L}_{d}$ and an amplitude spectrum loss $\mathcal{L}_{a}$. The data loss measures the difference between the NN's outputs $O_i$, $i=1, \cdot\cdot\cdot, N$ and the corresponding pseudo-labels $L_i$, $i=1, \cdot\cdot\cdot, N$, using the mean absolute error (MAE) as follows:
\begin{equation}\label{eq9}
\begin{gathered}
\mathcal{L}_{d}\left (L,O \right)=\frac{1}{N}\displaystyle \sum^{N}_{i=1}{\left|L_{i}-O_{i} \right|}.
\end{gathered}
\end{equation}
The amplitude spectrum loss $\mathcal{L}_{a}$ is obtained by computing the misfit in amplitude spectra between the network's output and the pseudo-labels, also employing the MAE metric as follows:
\begin{equation}\label{eq10}
\setlength{\abovedisplayskip}{5pt}
\setlength{\belowdisplayskip}{5pt}
\begin{gathered}
\mathcal{L}_{a}\left (L,O \right)=\frac{1}{N}\displaystyle \sum^{N}_{i=1}{\left|\text{AS}(L_{i})-\text{AS}(O_{i}) \right|},
\end{gathered}
\end{equation}
where the symbol $\text{AS}(\cdot)$ represents the operation for obtaining the amplitude spectrum. 

The total loss function is defined as
\begin{equation}\label{eq11}
\setlength{\abovedisplayskip}{5pt}
\setlength{\belowdisplayskip}{5pt}
\begin{gathered}
\mathcal{L}\left (L,O \right) = \mathcal{L}_{d}\left (L,O \right) + \epsilon \cdot \mathcal{L}_{a}\left (L,O \right),
\end{gathered}
\end{equation}
where $\epsilon$ is a hyperparameter, which is used to regulate the proportion of the amplitude spectrum loss within the total loss. In our numerical examples, we set it to a constant value of 0.01. The role of loss $\mathcal{L}_{a}$ and the setting of the hyperparameter $\epsilon$ will be thoroughly analyzed in the discussion section. We utilize the AdamW optimization algorithm \cite{loshchilov2017decoupled} in our traning stage. The implementation employs a GeForce RTX 8000 graphics processing unit and leverages the PyTorch framework. 

\section{Numerical examples}
\subsection{Synthetic Data}
\begin{figure*}[!t]
\centering
\includegraphics[width=0.5\textwidth]{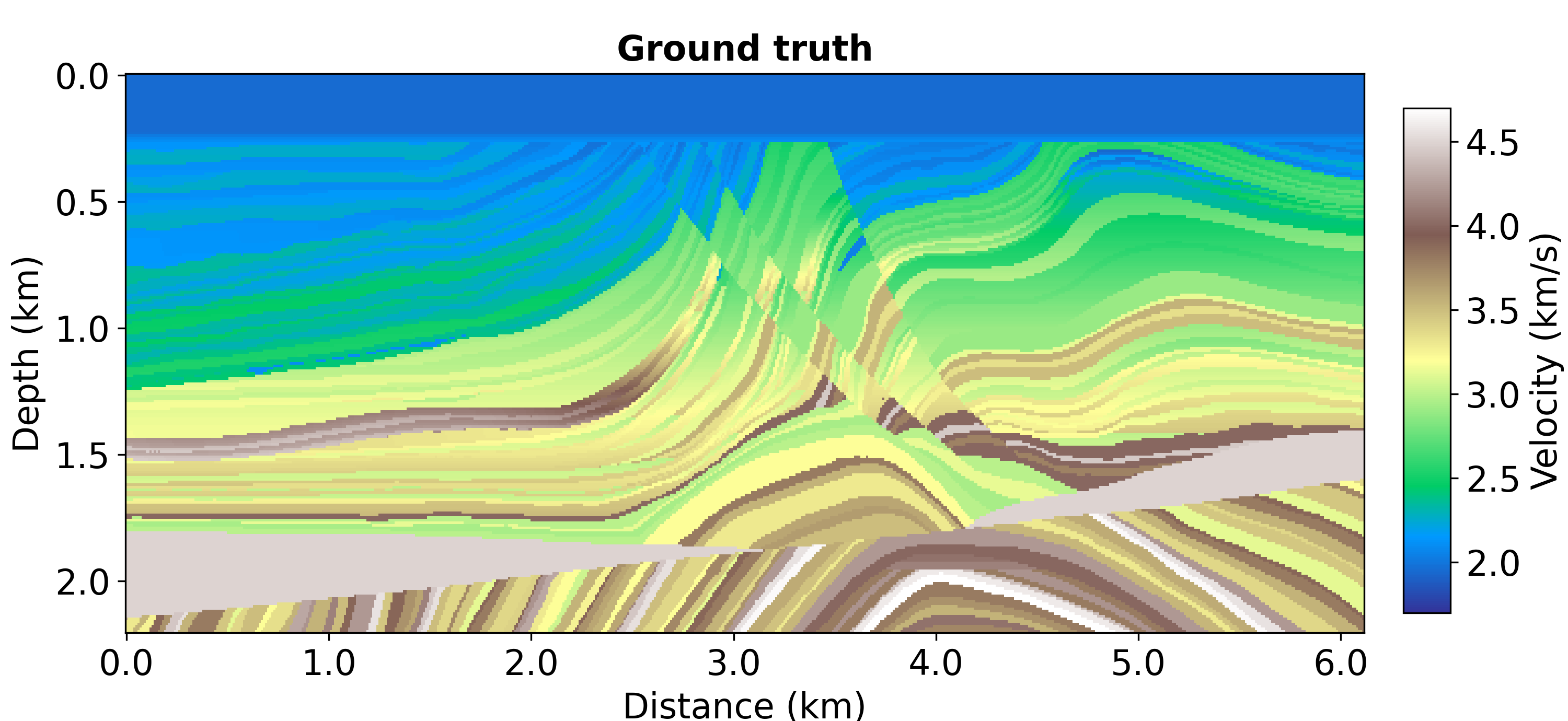}
\caption{True Marmousi velocity model. }
\label{fig3}
\end{figure*}

In order to verify the reasonability and accuracy of the proposed algorithm, we first use the simulated data generated with Marmousi model to perform the following tests. The modified model size is $2.20 \text{km} \times 6.11 \text{km}$ with a spatial interval of $10.0 \text{m}$ along both horizontal and vertical directions. Figure \ref{fig3} displays the true Marmousi velocity model with the modified dimensions. A Ricker wavelet with dominant frequency of 15 Hz is used as the source signal to generate the mimic observed seismic data. We generate a total of 122 shot gathers, each recording 2401 time steps with a time interval of 0.002s. The distance between shots is 50m, with the initial shot located at the model's far left at zero km. From these 122 shot gathers, we extract a total of 8200 data patches, each sized 128x128. For each patch we randomly choose the cutoff low-frequencies between $5\sim15$ Hz, forming our original observed dataset. During the warm-up phase, we randomly filter out frequency components below $6\sim30$ Hz from the original observed data to generate the input data. As previously described, the pseudo-label data at this time is the original observed data. In the IDR phase, the high-pass filter cutoff frequency range for the initial 50 epochs is set between $5\sim7$ Hz. Then, every 50 epochs, we increase the maximum cutoff frequency by 2 Hz, while the minimum cutoff frequency remains at 5 Hz. Once the maximum cutoff frequency reaches 15 Hz, it is kept fixed. The network is trained for a total of 550 epochs, with the warm-up phase accounting for 50 epochs. The learning rate start with 3e-4, then decreased by a factor of 0.8 at the 65, 130, 195, 260, and 325 epochs.

\begin{figure*}[!t]
\centering
\includegraphics[width=1.0\textwidth]{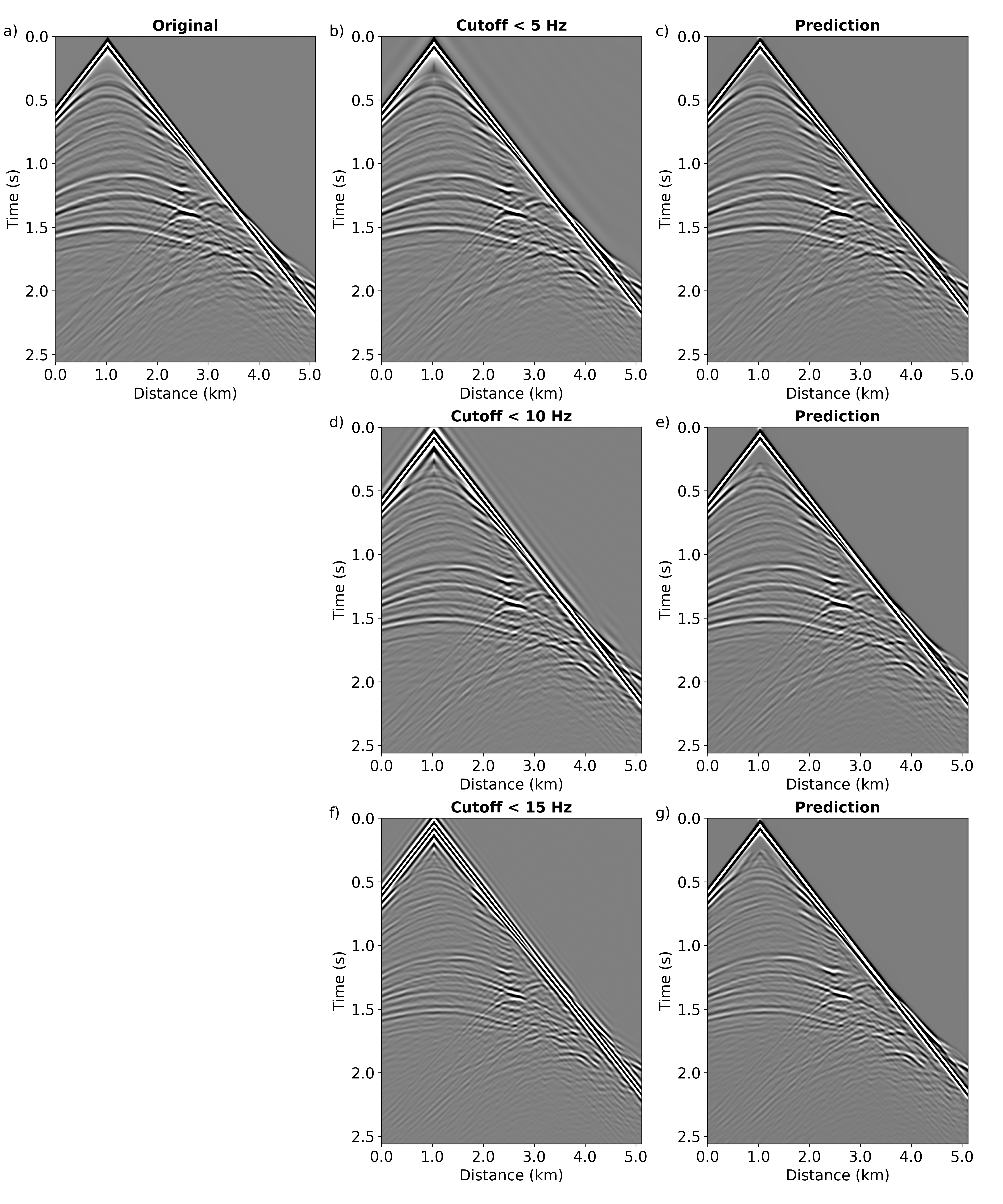}
\caption{The extrapolation results for the test shot gathers missing different low-frequency components. (a) the original full band data; data filtered by different high-pass filters: (b) 5 Hz, (d) 10 Hz and (f) 15 Hz; the corresponding data recovered with the proposed SSL algorithm: (c) 5 Hz, (e) 10 Hz and (g) 15 Hz.}
\label{fig4}
\end{figure*}

Figure \ref{fig4}a shows the seismogram of a simulated shot gather. We use a high-pass filter to cut off different frequency components, as shown in Figures \ref{fig4}b, \ref{fig4}d and \ref{fig4}f filtered with 5 Hz, 10 Hz, 15 Hz high-pass filters, respectively. After processing with our trained network, we can recover the missing components and the corresponding recovered data are shown in Figures \ref{fig4}c, \ref{fig4}e and \ref{fig4}g.

\begin{figure*}[!t]
\centering
\includegraphics[width=1.0\textwidth]{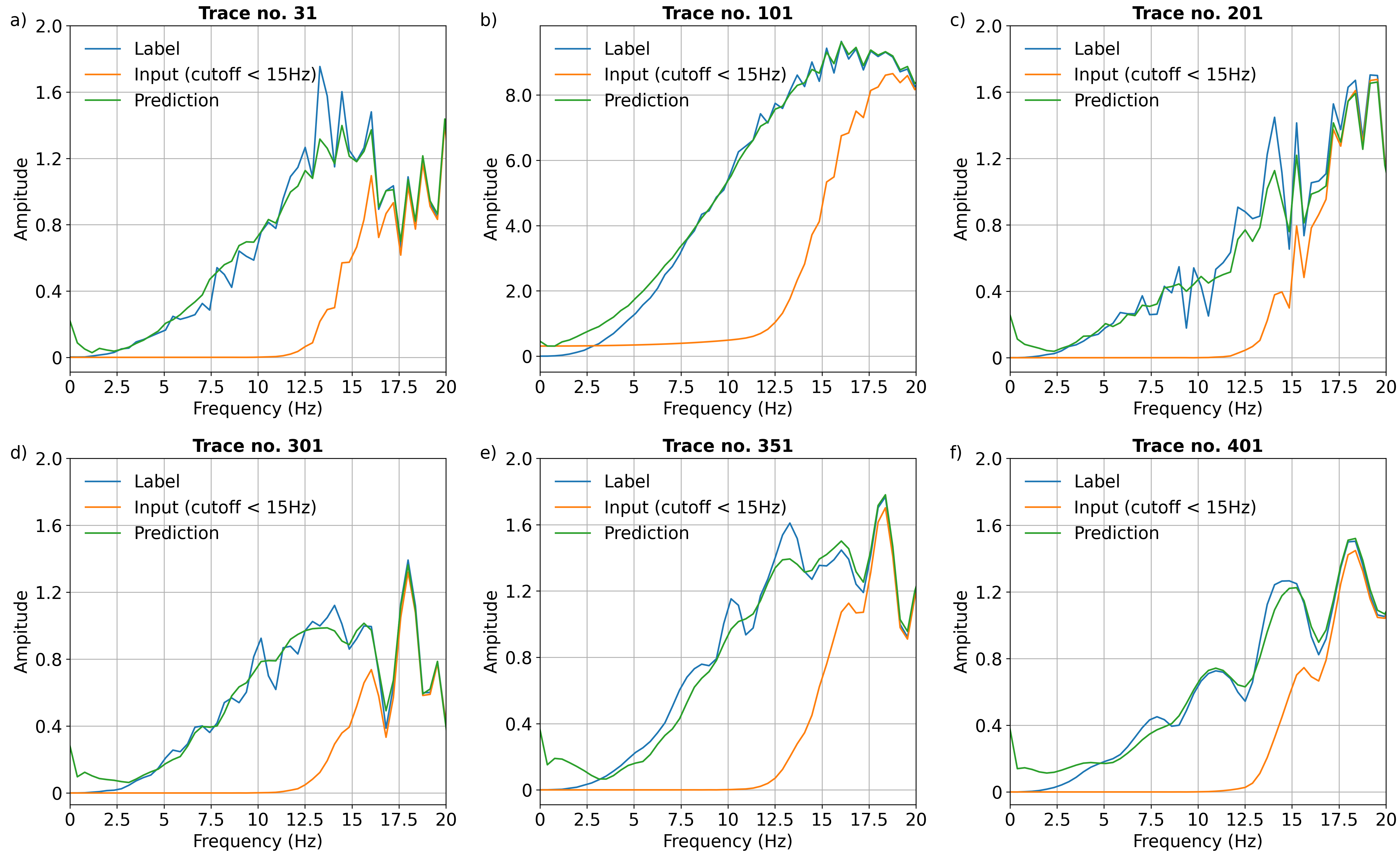}
\caption{The amplitude spectrum comparison at different locations. The 10 Hz high-pass filter is used. }
\label{fig5}
\end{figure*} 

\begin{figure*}[!t]
\centering
\includegraphics[width=1.0\textwidth]{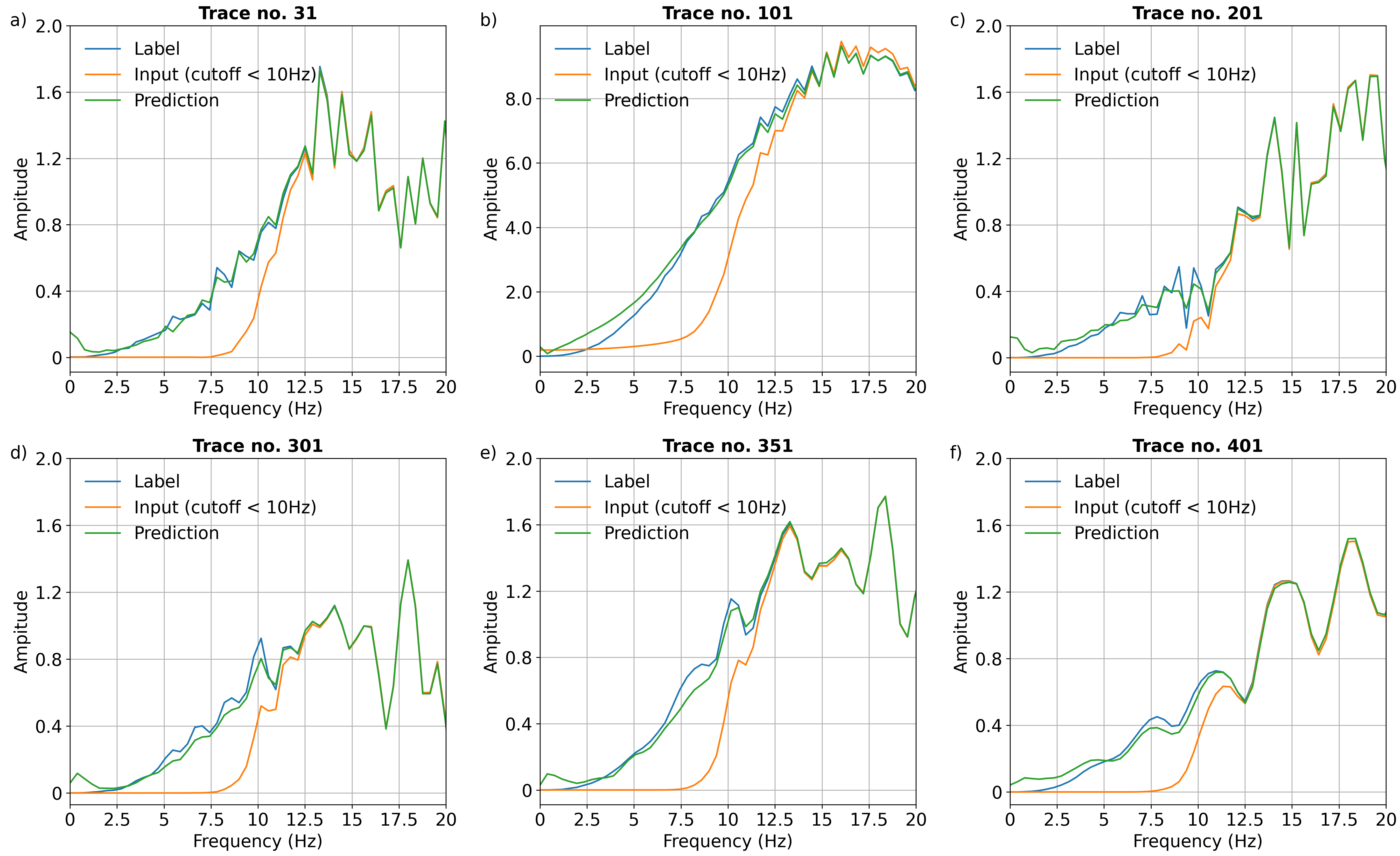}
\caption{The amplitude spectrum comparison at different locations. The 15 Hz high-pass filter is used.  }
\label{fig6}
\end{figure*} 

\begin{figure*}[!t]
\centering
\includegraphics[width=1.0\textwidth]{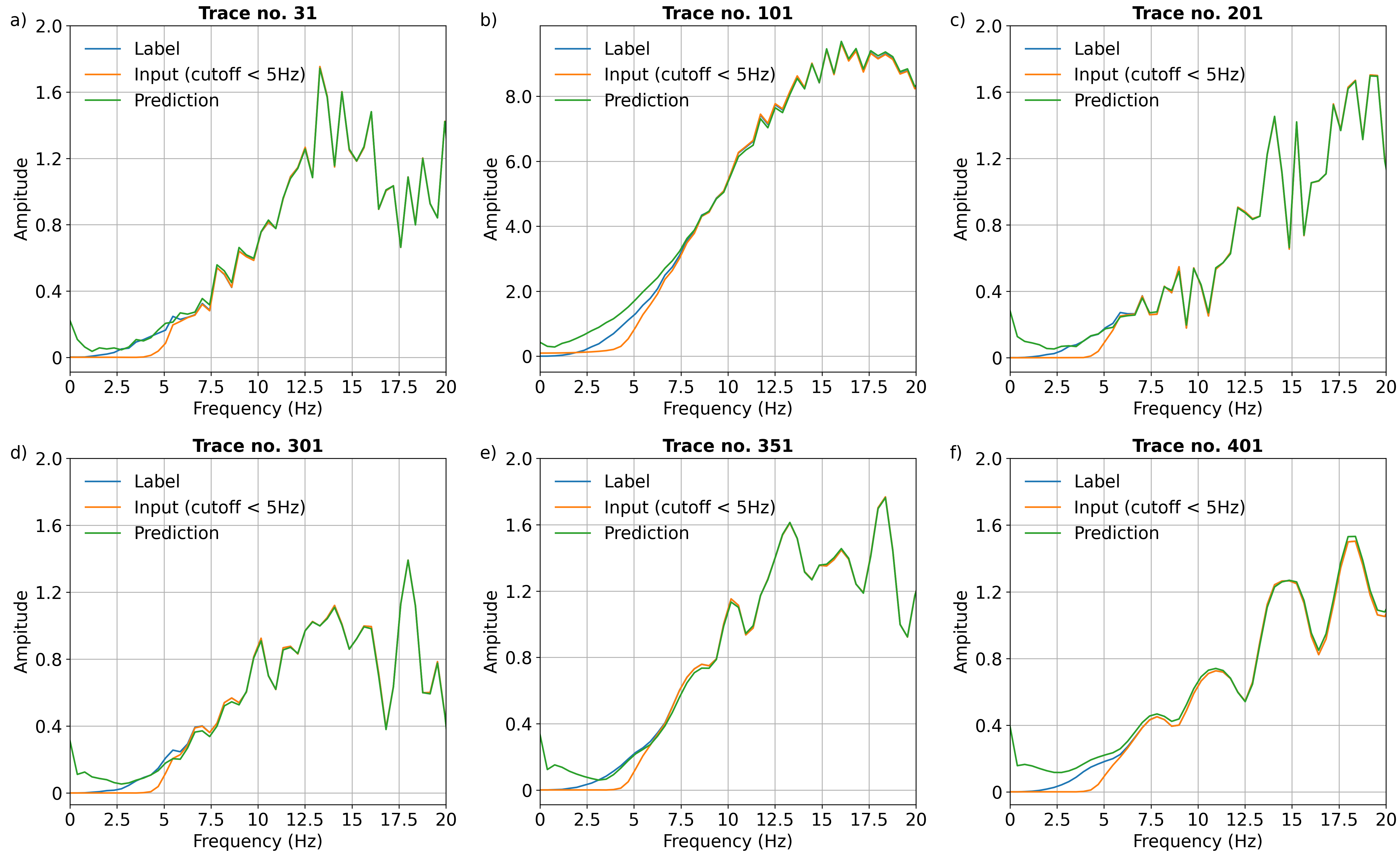}
\caption{The amplitude spectrum comparison at different locations. The 5 Hz high-pass filter is used.  }
\label{fig7}
\end{figure*} 

\begin{figure*}[!t]
\centering
\includegraphics[width=1.0\textwidth]{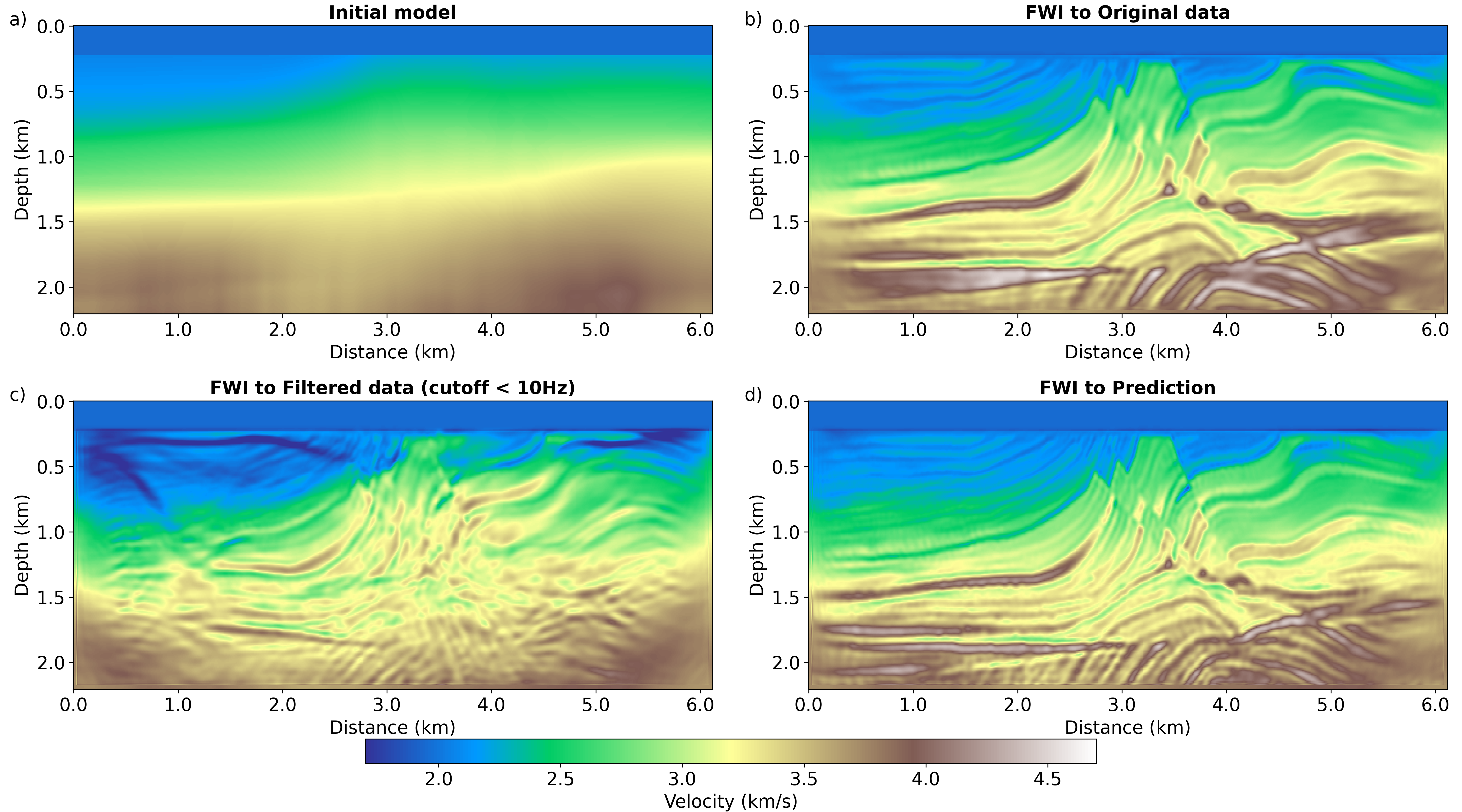}
\caption{(a) The initial velocity model and inversion results generated with different data sets: (b) original full band data, (c) filtered data with a cutoff frequency of 10 Hz and (d) the predicted data with the proposed algorithm. }
\label{fig8}
\end{figure*} 

\begin{figure*}[!t]
\centering
\includegraphics[width=1.0\textwidth]{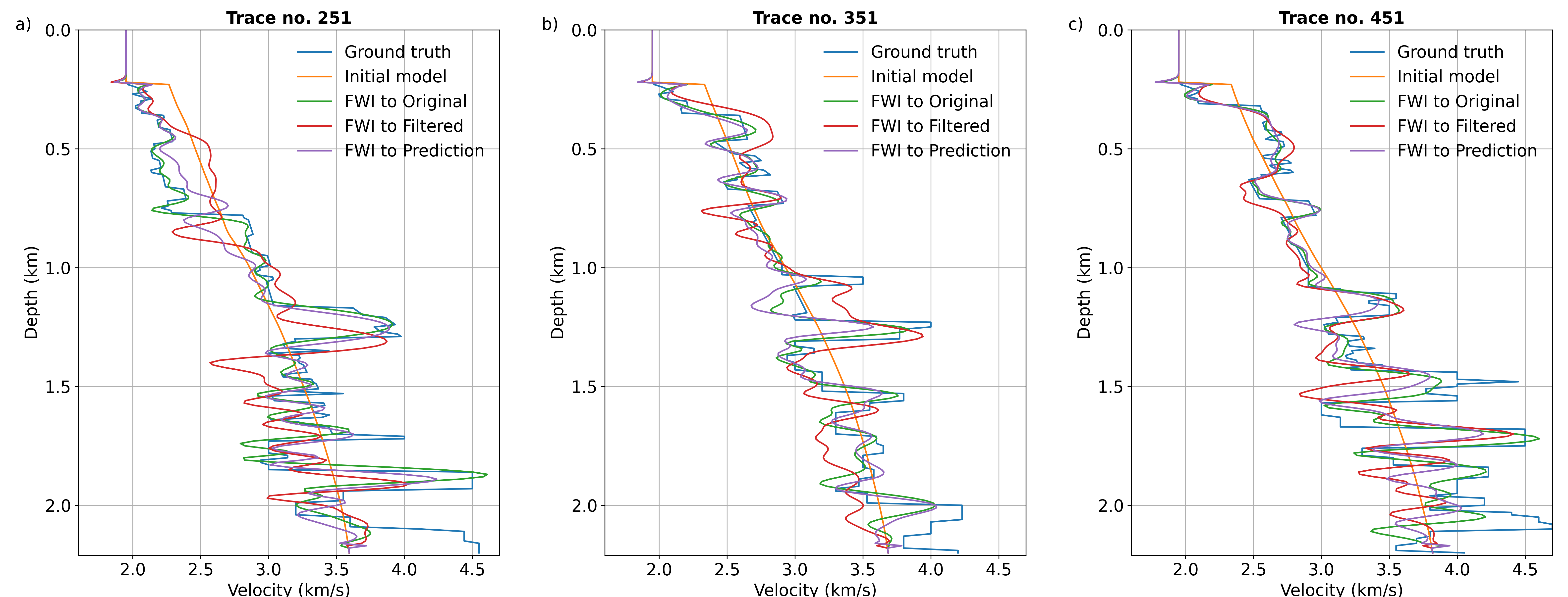}
\caption{Profiles at different locations: (a) X=2.5 km, (b) X=3.5 km and (c) X=4.5 km. The cutoff frequency is 10 Hz. }
\label{fig9}
\end{figure*} 

\begin{figure*}[!t]
\centering
\includegraphics[width=1.0\textwidth]{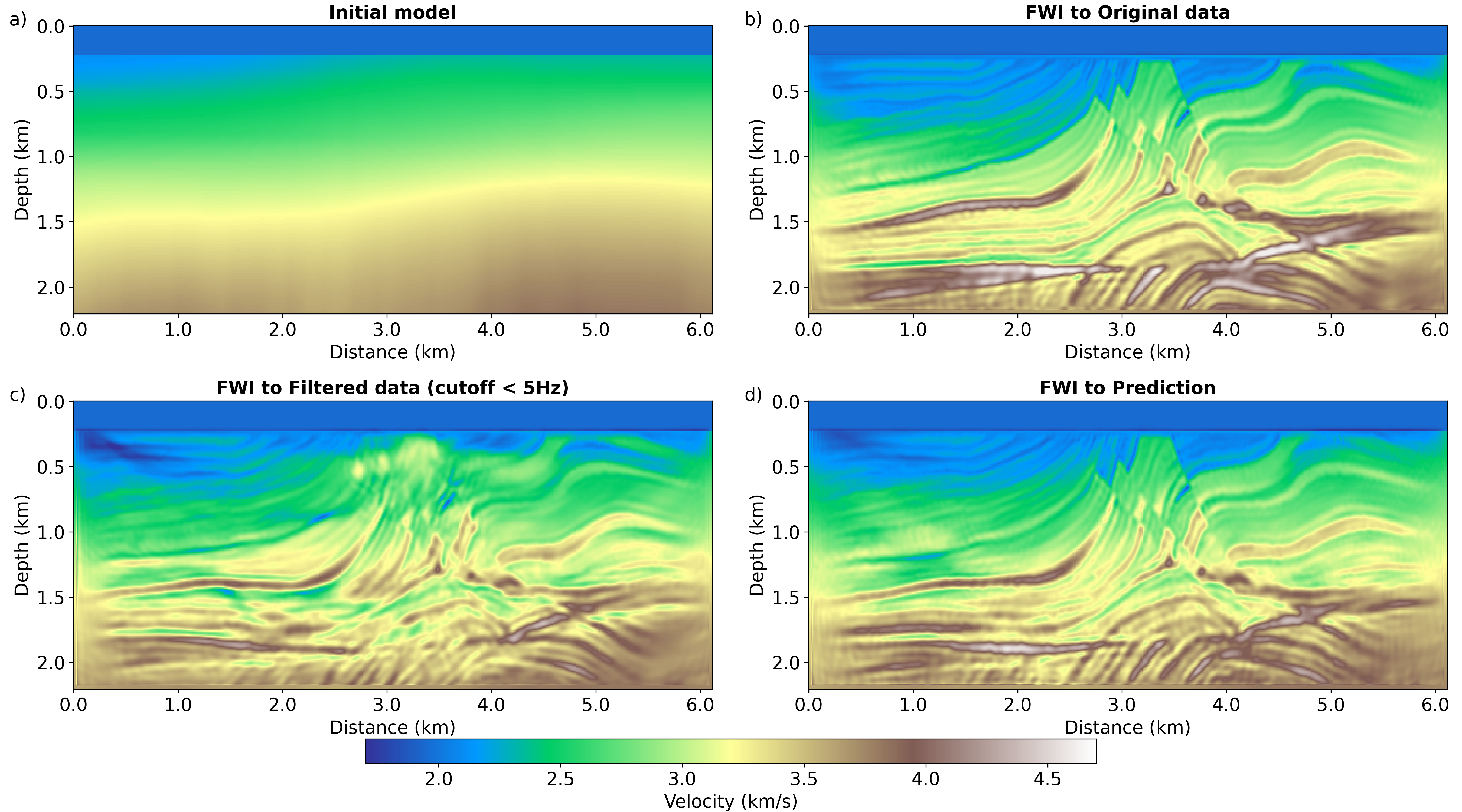}
\caption{(a) The initial velocity model and inversion results generated with different data sets: (b) original full band data, (c) filtered data with a cutoff frequency of 5 Hz and (d) the predicted data with the proposed algorithm. }
\label{fig10}
\end{figure*} 

\begin{figure*}[!t]
\centering
\includegraphics[width=1.0\textwidth]{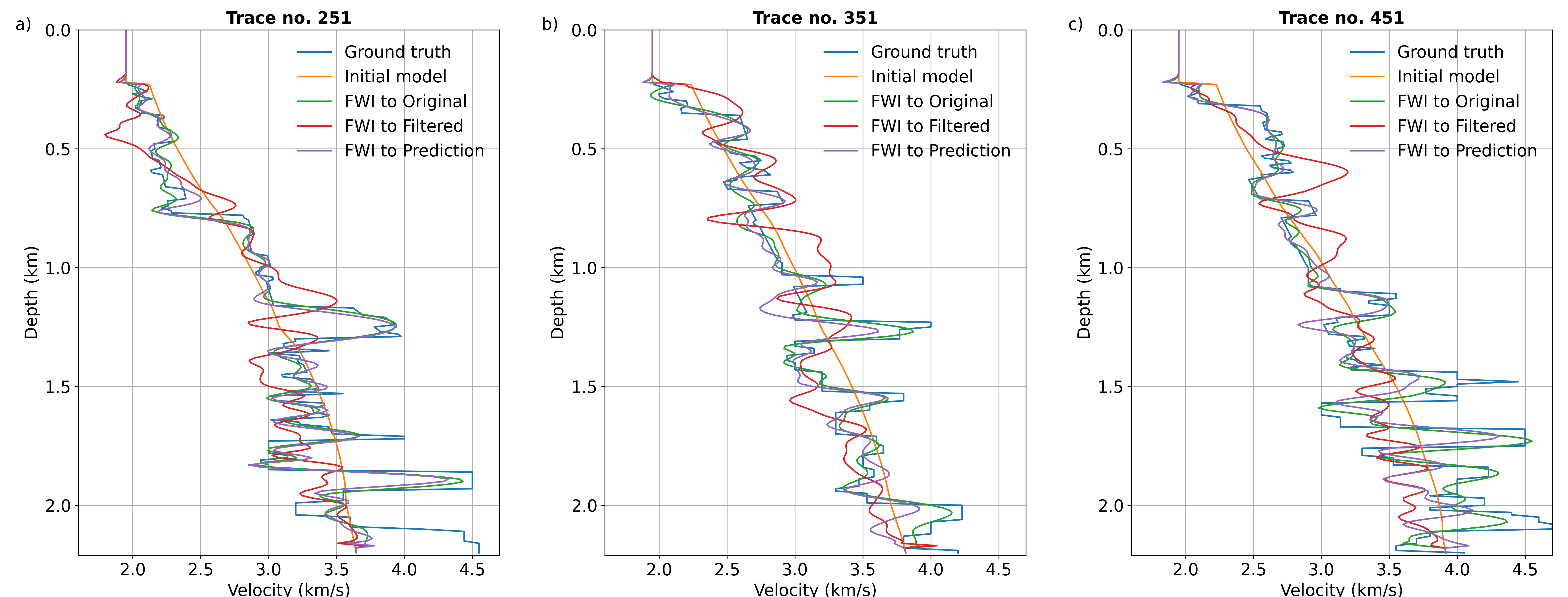}
\caption{Profiles at different locations: (a) X=2.5 km, (b) X=3.5 km and (c) X=4.5 km. The cutoff frequency is 5 Hz. }
\label{fig11}
\end{figure*} 

\begin{figure*}[!t]
\centering
\includegraphics[width=0.85\textwidth]{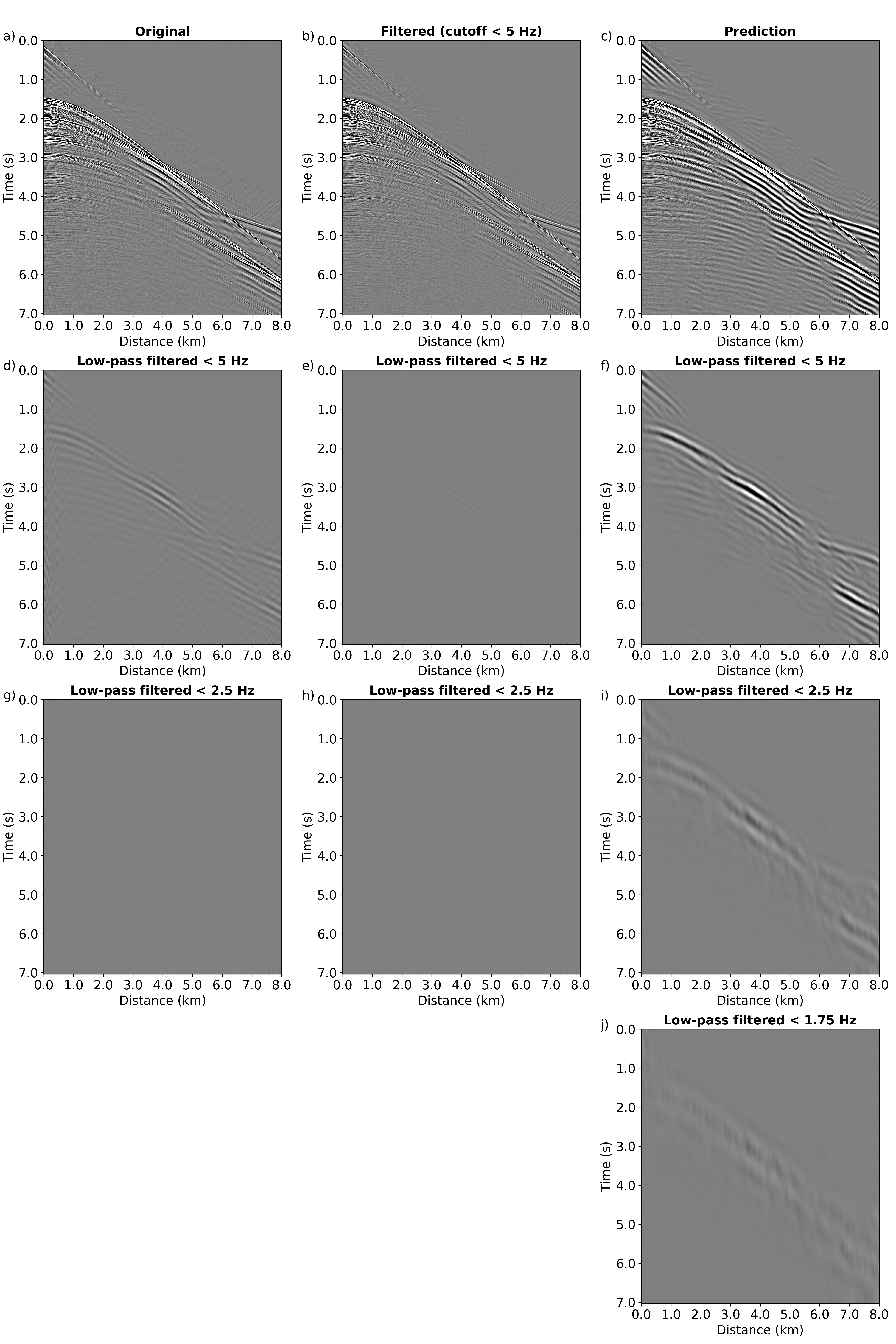}
\caption{Real marine seismic data, which were acquired from North West Australia. (a) the original observed data, (b) the data filtered by 5 Hz high-pass filters, (c) the predicted data with the proposed algorithm. The frequency components less than 5 Hz: (d) original data, (e) filtered data and (f) predicted data. The frequency components less than 2.5 Hz: (g) original data, (h) filtered data and (i) predicted data. The frequency components less than 1.75 Hz: (j) predicted data.}
\label{fig12}
\end{figure*} 

\begin{figure*}[!t]
\centering
\includegraphics[width=1.0\textwidth]{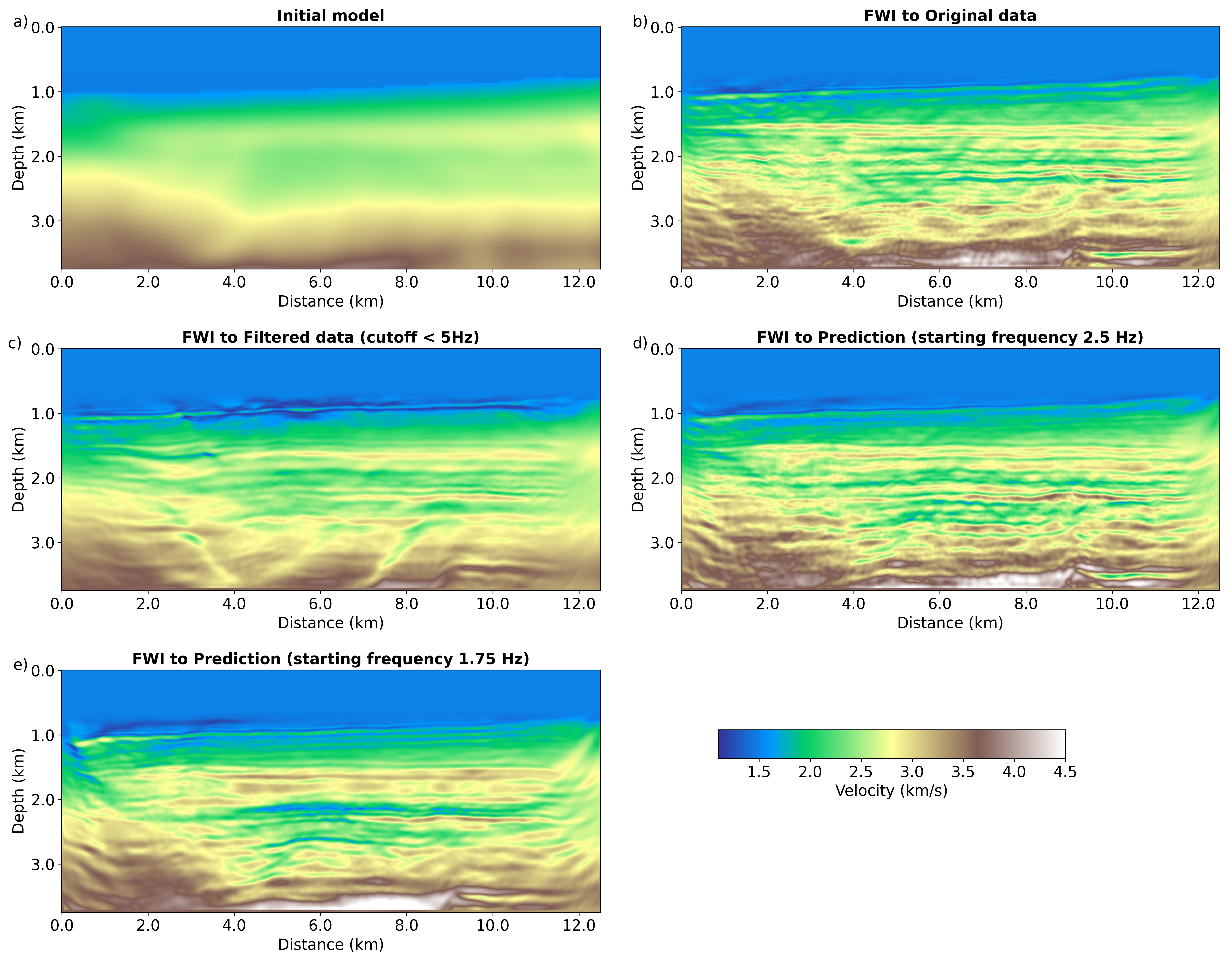}
\caption{(a) The initial velocity model. Inversion results of different data sets: (b) the original observed data, (c) the data with cutoff frequency of 5 Hz, (d) the predicted data (with starting frequency of 2.5 Hz), (e) the predicted data (with starting frequency of 1.75 Hz).}
\label{fig13}
\end{figure*} 

\begin{figure*}[!t]
\centering
\includegraphics[width=1.0\textwidth]{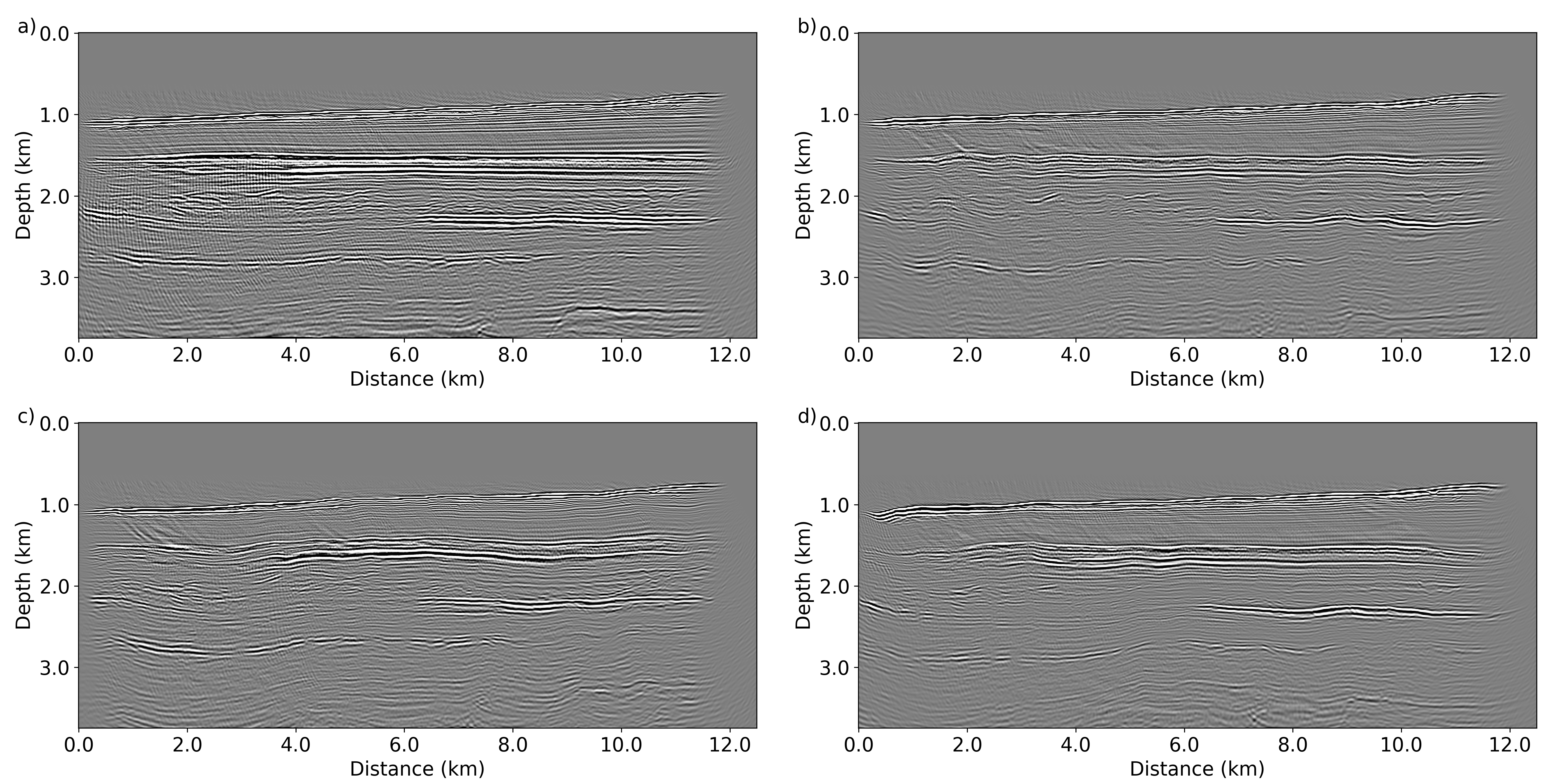}
\caption{Reverse time migration images generated with different velocity models: (a) the initial velocity model, (b) the inversion result of original observed data, (c) the inversion result of data with cutoff frequency of 5 Hz, (d) the inversion result of the predicted data (with starting frequency of 1.75 Hz).}
\label{fig14}
\end{figure*} 

\begin{figure*}[!t]
\centering
\includegraphics[width=1.0\textwidth]{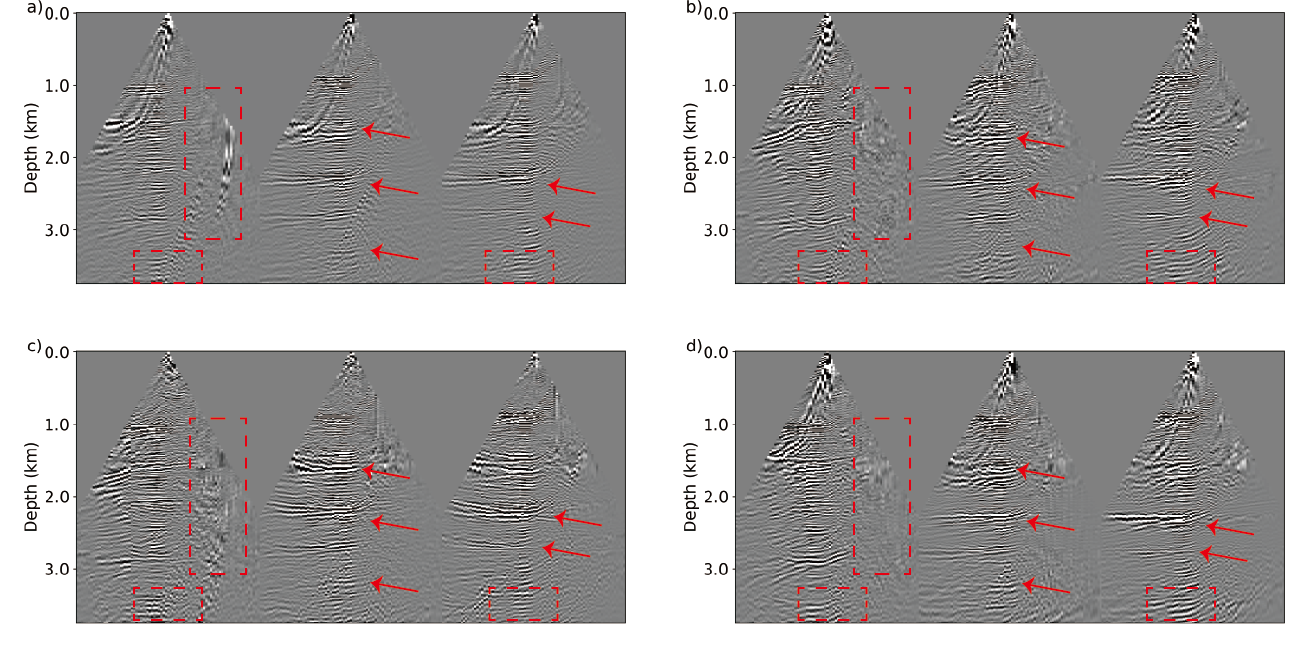}
\caption{Common image gathers at X=2.15 km, 5.5 km and 6.9 km using: (a) the initial velocity model, (b) the inversion result of original observed data, (c) the inversion result of data with cutoff frequency of 5 Hz, (d) the inversion result of the predicted data (with starting frequency of 1.75 Hz).}
\label{fig15}
\end{figure*}

To clearly verify the performance of the proposed algorithm in reconstructing the low-frequency components, we perform a spectral analysis of the traces at different locations between the original, filtered and predicted data, as displayed in Figures \ref{fig5}-\ref{fig7}. The predicted amplitude spectrum of data with cutoff frequency of 15 Hz (Figures \ref{fig5}) and 10 Hz (Figure \ref{fig6}) exhibit a good recovery of the filtered out components after applying the proposed algorithm. The reconstructed low-frequency components (green line) commendably match the original true information (orange line) at the missing band range. As for the more practical case with cutoff frequency of 5 Hz (Figure \ref{fig7}), the proposed extrapolation algorithm can still recover the missing low-frequency information even to 1.5 Hz. Therefore, Figures \ref{fig5}-\ref{fig7} prove the accuracy, stability and reasonability of the proposed SSL low-frequency extrapolation algorithm.

With the predicted data after low-frequency extrapolation, we perform the following FWI tests. Figure \ref{fig8} shows the inversion results with  cutoff frequency of 10 Hz. Figure \ref{fig8}a displays the smoothed initial velocity model, which is far away from the true model in Figure \ref{fig3}. We choose 5 frequency bands with dominant frequencies of 3 Hz, 4 Hz, 5 Hz, 8 Hz and 15 Hz to perform the multi-scale inversion, Figure \ref{fig8}b is the corresponding result of the original full-band data, matching the true model perfectly. If we directly use the data with cutoff frequency of 10.0 Hz to conduct the test, the inversion result is contaminated with artifacts because of the missing low-frequency information (see Figure \ref{fig8}c). In comparison, we can reconstruct the subsurface velocity model well with the predicted seismic data (see Figure \ref{fig8}d). To precisely show the inversion results, we choose the profiles at X=2.5 km, 3.5 km and 4.5 km and exhibit them in Figure \ref{fig9}. We can see that the direct inversion of filtered data (red line) is far away from the true model (blue line). Encouragingly, the inversion result of the predicted data (purple line) is similar to that of the full-band original data (green line), both of which match the true model perfectly.

In line with real seismic exploration scenarios, we also perform the tests on the data missing frequency information below 5 Hz. In order to highlight the importance of the low frequency components, we use a smoother velocity as the initial model (Figure \ref{fig10}a). Five frequency bands with dominant frequencies of 2 Hz, 3 Hz, 5 Hz, 8 Hz and 15 Hz are selected to conduct the multi-scale inversion. Figure \ref{fig10}b displays the result of the inversion applied to the original full-band data, which also nicely matches the true model. When using the filtered data missing frequency information below 5 Hz to conduct the test, the inversion result is poor due to the cycle-skipping issue. With the proposed algorithm to extrapolate the filtered data, the inversion result can recover the velocity well. Figure \ref{fig11} displays the profiles at X=2.5 km, 3.5 km and 4.5 km, from which we can see that the result of the filtered data has both numerical and depth errors compared to the true model. In contrast, the result of using the predicted data fits the true model very well similar to the inversion applied to the original data.

\subsection{Field Data}

After verifying the accuracy and reasonability of the proposed algorithm, we attempt to apply it to real marine seismic data. The data were acquired in North West Australia using a variable depth streamer. The original dataset comprises 1824 shots spaced approximately 18.75 m apart laterally. For our test, we select 201 shots, with 56m lateral spacing, to lower the computational burden. Each shot gather is acquired through steamer cables with a maximum offset of 8.0 km and a 12.5 m recording spacing. 

The original acquired data included frequency components as low as 2.5 Hz. In our experiment, to consider a more general scenario where typically collected data lack frequencies below 5 Hz, we filtered out frequencies below 5 Hz from the original acquired data, treating it as our available observed data. We extract 18500 data patches from the filtered shot gathers, each sized 128x128. During the warm-up phase, the network undergoes pre-training for 50 epochs, where the input data is prepared by performing a high-pass filter with a cutoff frequency of 10 Hz on the observed data.. In the IDR phase, the initial range for the high-pass filter cutoff frequency is set between $1\sim2$ Hz, and then increases by 0.5 Hz for both the upper and lower limits every 20 epochs. Eventually, the filter cutoff frequency is fixed between $4\sim5$ Hz. Instead of starting from scratch, here the network is initialized from the model trained on synthetic data. The network is trained with a total of 200 epochs. The initial learning rate is 2e-4, which is reduced by a factor of 0.8 at the 25, 50, 75, 100, 125, and 150 epochs. 

Figure \ref{fig12}a denotes the original seismogram of the first shot. Figure \ref{fig12}b shows the corresponding filtered shot gather, where we cut off the information below 5 Hz. Figure \ref{fig12}c is the result after low-frequency extrapolation with the proposed algorithm. Figure \ref{fig12}d-\ref{fig12}f displays the energy below 5 Hz of the original, filtered and predicted data, from which we find that the original signal below 5 Hz has weaker energy than the predicted one. This is because the prediction does not have limitations in the low frequency limit, whereas the data contain limited energy below 2.5 Hz. As shown in Figures \ref{fig12}g-\ref{fig12}i, we manage to predicted low frequency information below 2.5 Hz. Figure \ref{fig12}j shows that there still exists some signal of reflected waves, which means that the proposed algorithm could extrapolate to frequencies as low as 1.75 Hz.

Figure \ref{fig13}a displays the initial velocity model, Figures \ref{fig13}b and \ref{fig13}d are the inversion results of the original and predicted data with starting dominant frequency of 2.5 Hz. The corresponding inversion results are similar to a large degree, indirectly proving that the frequency components within $2.5\sim5.0$ Hz predicted by our algorithm are consistent with the original data. However, the result of predicted data shows some additional details, which correspond to the additional components less than 2.5 Hz after the frequency extrapolation. If we directly use the filtered data to perform the inversion, the corresponding inversion result is shown in Figure \ref{fig13}c. We can see that two big false faults appear at X=3 km and 8 km caused by the cycle-skipping problem. To emphasize the positive performance of our proposed low-frequency extrapolation algorithm, we conduct a comparative inversion test with an initial dominant frequency of 1.75 Hz. The corresponding outcome is presented in Figure \ref{fig13}e. Contrasting the outcomes displayed in Figures \ref{fig13}b and \ref{fig13}d, the result depicted in Figure \ref{fig13}e reveals a more comprehensive representation of the background information.

To demonstrate the accuracy of the different velocity models shown in Figure \ref{fig13}, we perform a reverse time migration. Figure \ref{fig14} shows the corresponding migration images. Compared to the image from the initial model (Figure \ref{fig14}a), those of the FWI results (Figure \ref{fig14}b-\ref{fig14}d),to some degree, show the stratigraphic relief. To evaluate the accuracy of the images, we choose three common image gathers at X=2.15 km, 5.5 km and 6.9 km to make a detailed analysis, as shown in Figure \ref{fig15}. Compared to the other images, that of the predicted data with starting dominant frequency of 1.75 Hz resulted in less noise (depicted by the dashed square area) and more flat gathers (indicated by the red arrows), proving the accuracy of the proposed algorithm.

\subsection{Application on earthquake seismogram data}
We finally select an earthquake seismogram data to demonstrate the generalizability and feasibility of our method for extending the frequency band to ultra-low frequency components for large-scale collected data. This earthquake seismogram data were collected in western Sichuan, China. The distribution of dense seismic stations and the structure of the collection area are shown in Figure 16. The waveforms used were recorded from September 2006 to July 2009.

In our experiment of low-frequency extrapolation of earthquake seismogram, we implement several modifications compared to processing in exploration data: 1. Instead of extracting two-dimensional data patches to construct datasets, we extract one-dimensional data, each comprising 128 time steps, directly from the acquired data; 2. Given that our training data are one-dimensional, we adapt our network architecture to include one-dimensional convolutional baselines; 3. In the initial training epochs, due to the network's instability, the predicted low-frequency components include some ultra-low-frequency numerical artifacts. Consequently, in IDR phase, frequencies below 0.1 Hz are filtered out from the network's predicted pseudo-labels to ensure the stability of the network training process. We select two teleseismic events, specifically events 51 and 53, from the collected earthquake seismogram. The original collected data contain frequencies below 0.25 Hz. To consider the scenario of missing low-frequencies, we applied a high-pass filter with a cutoff frequency of 0.25 Hz to the original data. The filtered data, assumed as known, are used to construct our training dataset. Notably, the original collected data are not included in our training set, as our approach operates on an SSL basis. During the training's warm-up and IDR phases, we employ a similar strategy, progressively increasing the high-pass filter cutoff frequency range, ultimately fixing it at 0.25 Hz. The network undergoes training for a total of 300 epochs. The initial learning rate 1e-4, and a scheduler reduces it by a factor of 0.8 at the 30, 60, 90, 120, 150, and 180 epochs.

We present the original seismic waveforms of different teleseismic events (Figure \ref{fig17} for event 20 and Figure \ref{fig18} for event 73), along with the waveforms filtered to exclude frequency components below 0.25 Hz, and compared these with the predicted waveforms under various low-pass filters. Due to the complexity of the coda wave following the direct P phase, we establish a variable time window to better demonstrate the restoration of low-frequency components in the predicted waveforms relative to the original recordings. This time window is indicated by the light green shaded areas in Figures \ref{fig17} and \ref{fig18} (panel (a)). Within this window, we extract the band-passed original recordings (black), filtered waveforms (blue), and predicted waveforms (red). We then compute their normalized cross-correlation coefficients (NCC) across different filtering bands (0.01-0.15 Hz, 0.01-0.2 Hz, 0.01-0.25 Hz, 0.01-0.3 Hz). The NCC values thus reflect the degree of fit from the onset of the P-wave to the later coda waves. As observed in panels (b) of Figures \ref{fig17} and \ref{fig18}, for different low-pass filters, the predicted waveforms exhibit a better restoration of the original low-frequency components (the red NCC curves are consistently above the blue ones). These results demonstrate the effectiveness of our method in the restoration of ultra-low-frequency content in earthquake waveform data.

\begin{figure*}[!t]
\centering
\includegraphics[width=0.6\textwidth]{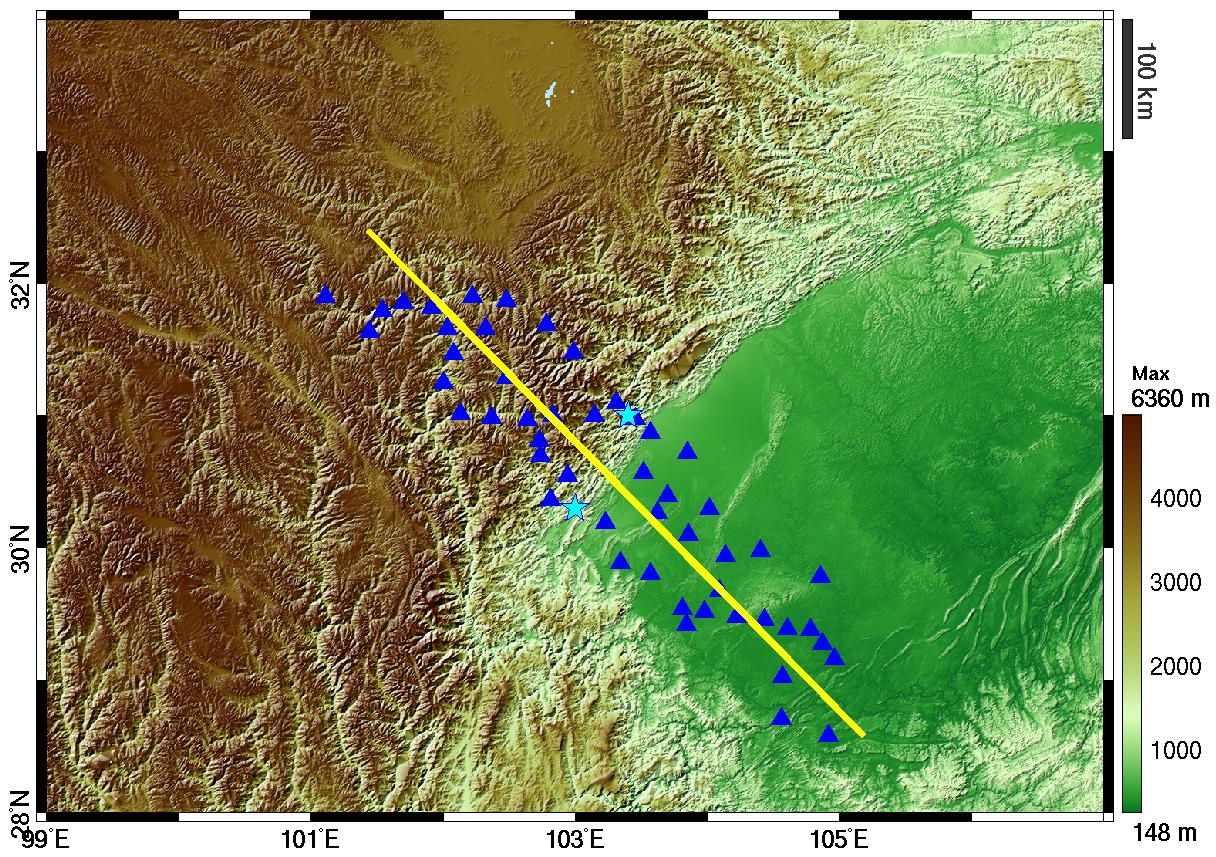}
\caption{The local structure and stations distribution in the selected western Sichuan for the study. The blue triangles in the figure represent the two-dimensional seismic array used in this research. All stations are arranged along a reflection profile: the Aba-Zigong measurement line (indicated by the thick yellow line). The epicenters of the 2008 Wenchuan earthquake (top right) and the 2013 Lushan earthquake (bottom left) are marked with cyan pentagrams.}
\label{fig16}
\end{figure*} 

\begin{figure*}[!t]
\centering
\includegraphics[width=0.33\textwidth]{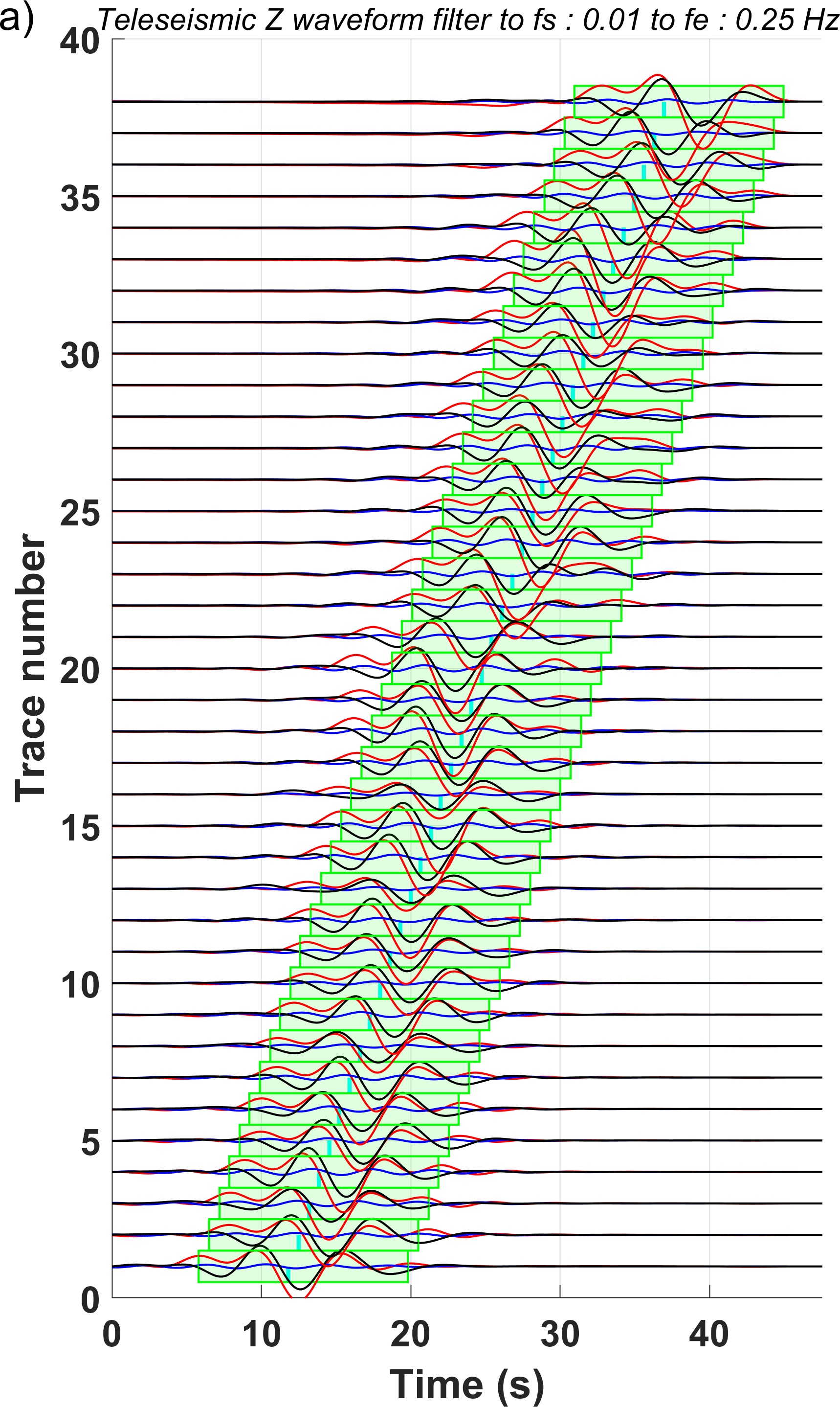}
\includegraphics[width=0.66\textwidth]{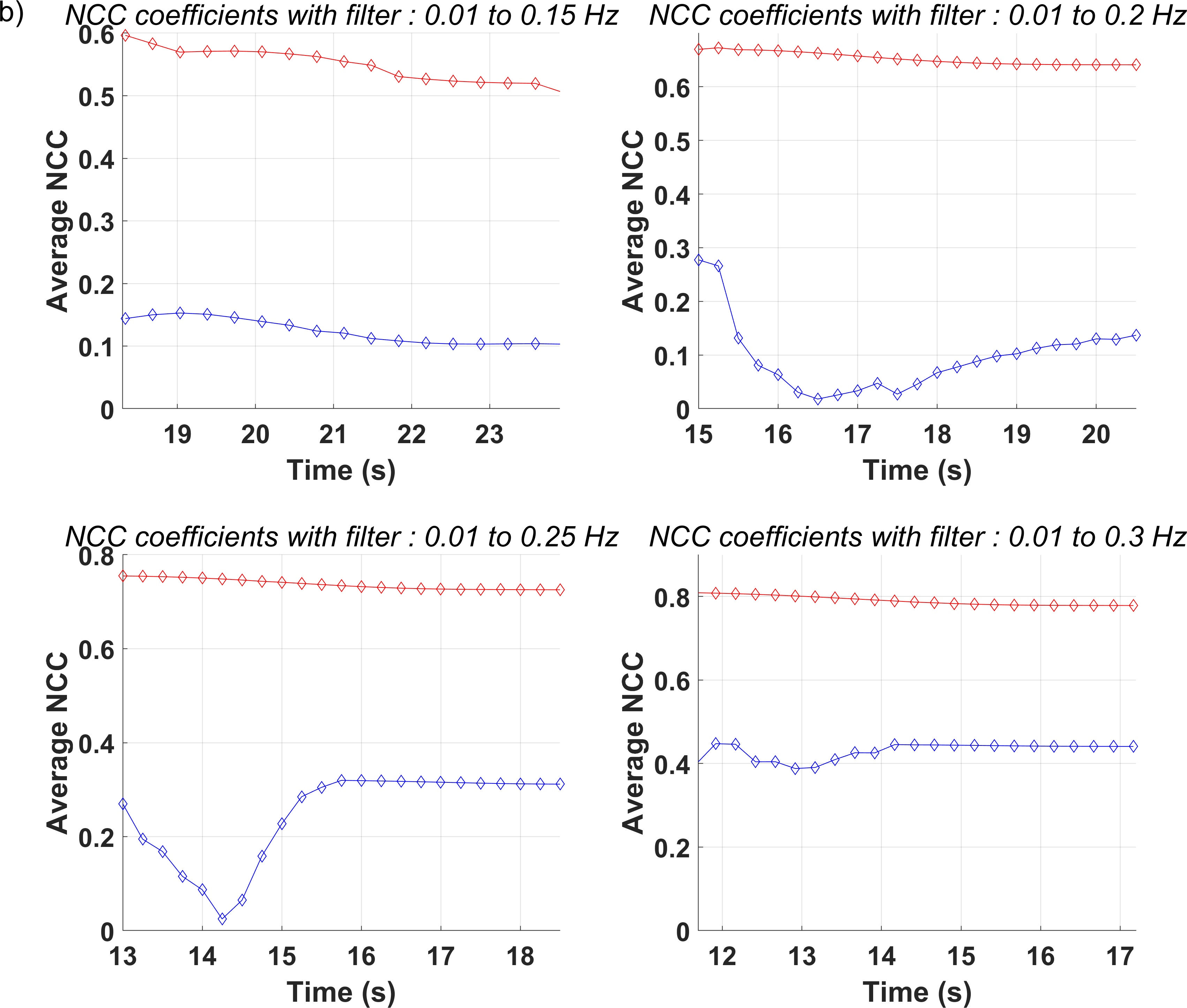}
\caption{The original recordings of the Z component received from teleseismic event 20, shown in black lines in panels (a), following a band-pass filter (0.01 to 0.25 Hz). The waveforms after filtering out frequencies below 0.25 Hz and then applying the same band-pass filter (0.01 to 0.25 Hz), illustrated in blue lines in panels (a). The predicted results to the filtered waveforms without low-frequency components below 0.25 Hz, followed by the same band-pass filter (0.01 to 0.25 Hz), represented by red lines in panels (a). The panel (b) illustrates a comparison of the normalized cross-correlation coefficients (NCC) for waveforms with and without low-frequency components below 0.25 Hz and predicted waveforms under four different band-pass filtering operations (from top left to bottom right: 0.01-0.15 Hz, 0.01-0.2 Hz, 0.01-0.25 Hz, 0.01-0.3 Hz). This comparison is conducted for different time windows. These windows are defined from two minimum periods (2/$f_e$, where $f_e$ is the highest cutoff frequency of the respective band-pass filter) before the maximum P-wave amplitude of the broadband waveform (indicated by the cyan dashed line) to varying lengths afterwards (ranging from 1s to a maximum of 2.25/$f_e$ s). The NCC coefficients are calculated from the band-pass filtered results of the original waveforms and either the predicted waveforms or those without low-frequency components below 0.25 Hz (represented by red and blue curves, corresponding to the colors in panels (a)).}
\label{fig17}
\end{figure*} 

\begin{figure*}[!t]
\centering
\includegraphics[width=0.33\textwidth]{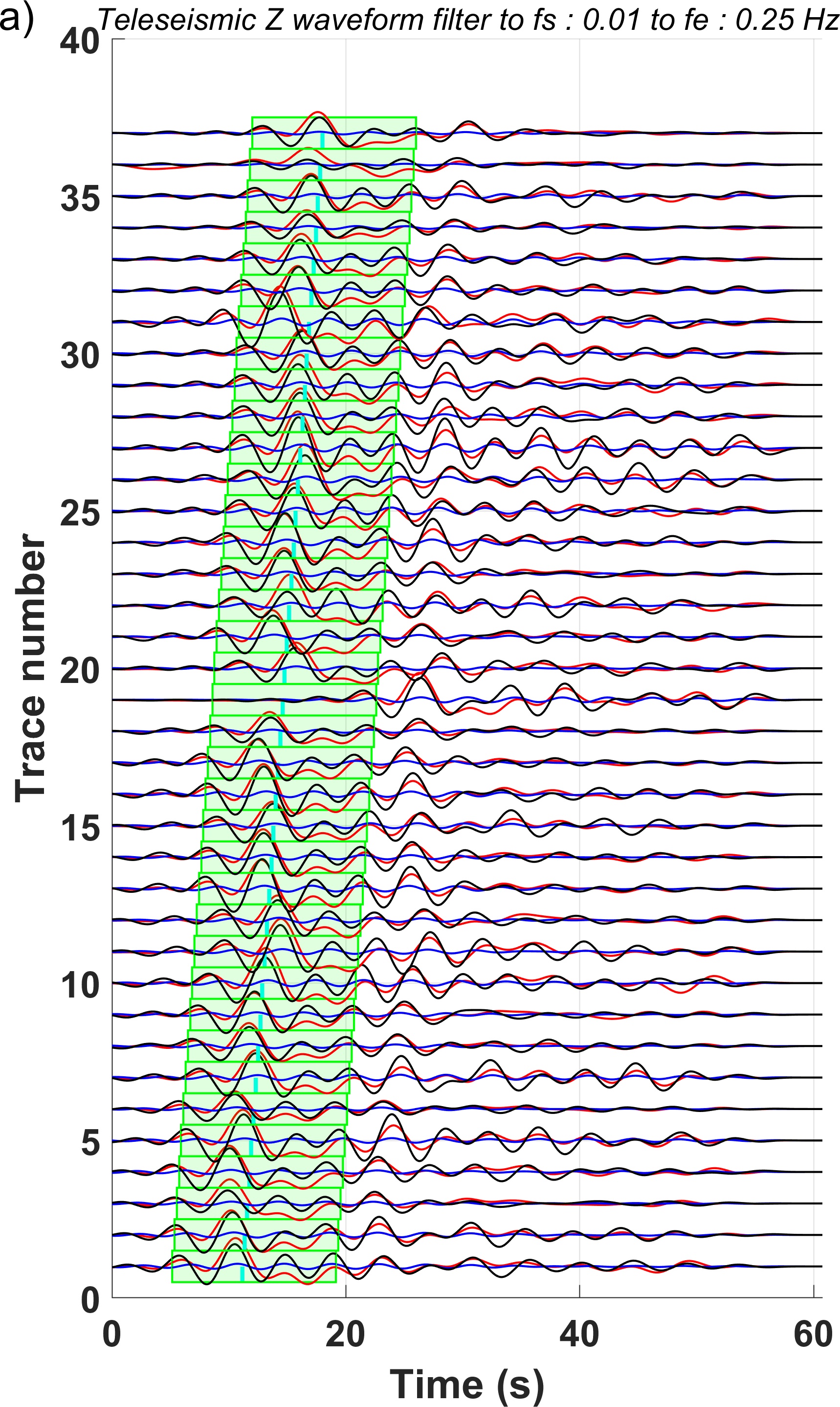}
\includegraphics[width=0.66\textwidth]{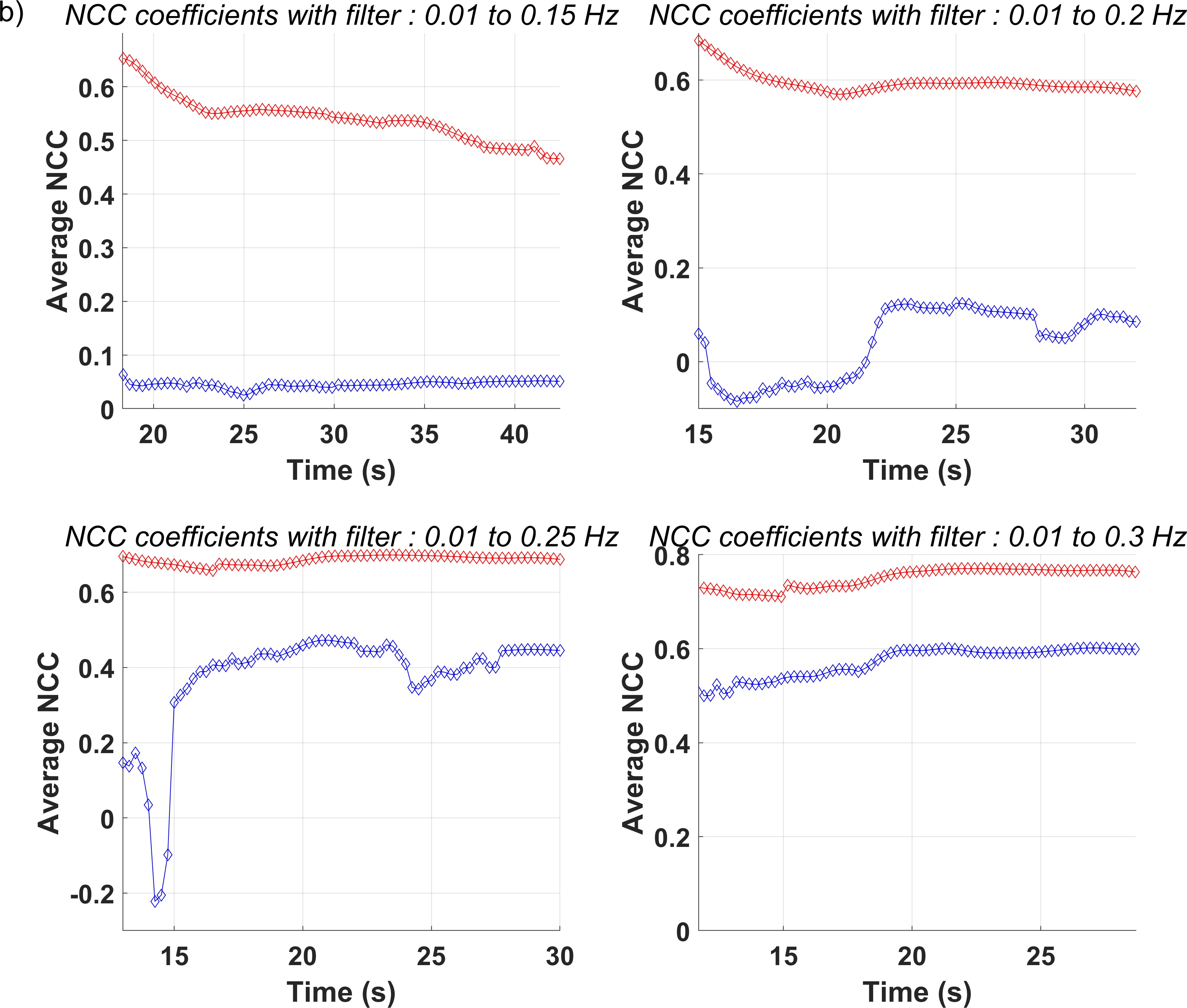}
\caption{The original recordings of the Z component received from teleseismic event 73, shown in black lines in panel (a), following a band-pass filter (0.01 to 0.25 Hz). The waveforms after filtering out frequencies below 0.25 Hz and then applying the same band-pass filter (0.01 to 0.25 Hz), illustrated in blue lines in panel (a). The predicted results of the filtered waveforms without low-frequency components below 0.25 Hz, followed by the same band-pass filter (0.01 to 0.25 Hz), represented by red lines in panels (a). The other figures carry explanatory notes consistent with Figure \ref{fig17}.}
\label{fig18}
\end{figure*}

\section{Discussion}
In this paper, we have concurrently validated the efficacy of our developed low-frequency extrapolation method on both exploration seismic and earthquake seismogram data. Owing to our method's adoption of a self-supervised learning (SSL) paradigm, it effectively obviates the need for labels, thereby enhancing its practical application potential. Subsequently, we will discuss the robustness of our method against noisy data, the role of the amplitude spectrum loss, and the strategies for the effective setting of filtering frequencies. 

\subsection{Robustness to noise}
It is widely recognized that observed seismic data are invariably contaminated by noise. Therefore, a critical examination is the robustness of our method against noise in the data. Recently, an SSL paradigm for noise reduction have proven its efficacy in attenuating various types of seismic noise \cite{cheng2023effective}. By integrating a noise reduction functionality into our SSL low-frequency extrapolation approach, we can present a unified SSL framework. This framework is capable of performing both denoising and low-frequency extrapolation processing on seismic data simultaneously. In the following, we will meticulously explore how to make fine-tuned adjustments to the developed SSL algorithms (Algorithm 1) to fulfill this purpose.

Initially, during the warm-up phase, we need to alter the generation mode of input data in the lesslow-low dataset. This entails adding noise to the original noisy data after it has been high-pass filtered, for example,
\begin{equation}\label{eq12}
I_i=H[x_i] + n_i, ~ i=1, \cdot \cdot \cdot, N,
\end{equation}
where the $I_i$ represents the input data, and $n_i$ denotes the added noise. The pseudo labels, however, remain the original noisy data $x_i$. It is important to note that the original noisy data also lack low frequency content. In other words, the training of our coupled framework for low-frequency extrapolation and denoising is conducted under an SSL paradigm.

Similarly, during the IDR stage, we create a new lesslow-low dataset's input by adding noise to the network's predicted results after high-pass filtering, such as
\begin{equation}\label{eq13}
I_i=H[\text{NN}(x_i)] + n_i, ~ i=1, \cdot \cdot \cdot, N.
\end{equation}
The pseudo labels are still the network's predictions of the original noisy data $\text{NN}(x_i)$. With these modifications, we can see that the input data, compared to the original noisy data, not only lack more low-frequency components but also contain a stronger noise. Consequently, the network is trained to optimize two objectives simultaneously: denoising and low-frequency extrapolation.

We share an example to validate the performance of the coupled SSL approach. We continue to employ the synthetic data generated in Section 3.1. The original noisy data exhibit a deficiency in the low-frequency range of $(5, 10) \text{Hz}$, and they include random noise with the levels ranging from 5 to 30, generated by the subsequent equation:
\begin{equation}\label{eq14}
n_i=0.01\cdot\xi \cdot std(L_i) \cdot rand(0,1), ~ i=1, \cdot \cdot \cdot, N,
\end{equation}
where the $\xi$ is the noise level, $std(L_i)$ represents the standard deviation of the pseudo label $L_i$, and $rand(0,1)$ is the standard normal distribution.

In the warm-up phase, the network is pre-trained for 50 epochs. When we create the lesslow-low dataset, the original noisy data are filtered within the range of $6\sim25 \text{Hz}$. The noise level added to the post-filtered results, ranging from 5 to 30, is used to construct the input data. In the IDR phase, the initial 50 epochs involve a random high-pass cutoff frequency ranging from 5 to 6 Hz. Subsequently, every 50 epochs, the upper limit of the high-pass cutoff frequency is increased by 1 Hz. This cutoff frequency range is fixed when it reaches the range of $5\sim10 \text{Hz}$. Throughout this entire phase, the level of noise added also varies randomly within the range of 5 to 30. The network in total undergoes training across 290 epochs. The initial learning rate is 2e-4, and a scheduler reduces it by a factor of 0.8 for every 65 epochs. 

\begin{figure*}[!t]
\centering
\includegraphics[width=1.0\textwidth]{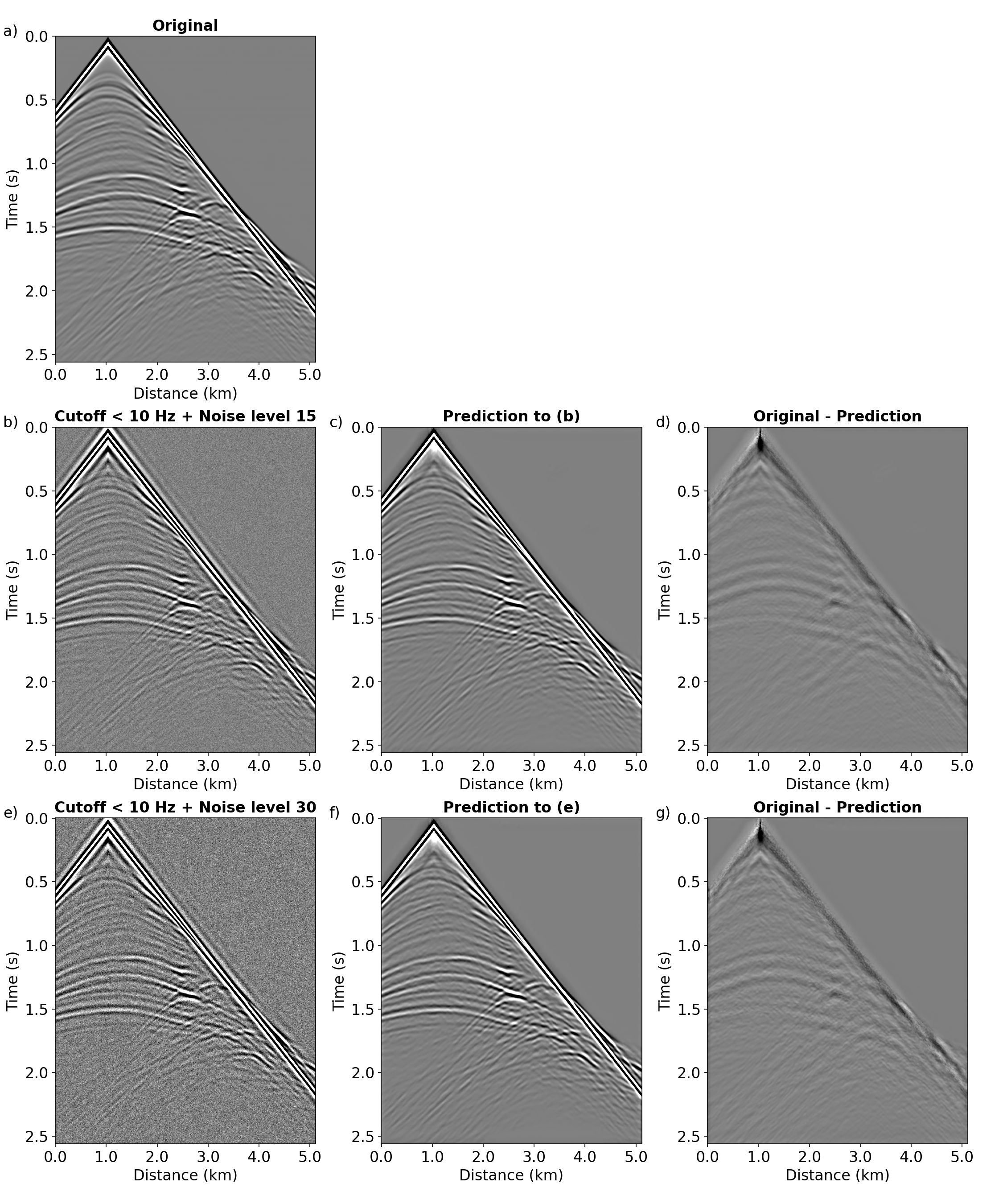}
\caption{Testing the capability of our method to simultaneously restore low-frequency components and denoise the noisy data with missing low frequencies: (a) the mimic observed seismogram generated with the true velocity model, (b) and (e) are the test data missing low frequencies below 10 Hz but containing noise with a level of 15 and 30, respectively, (c) and (f) correspond the network's prediction results to (b) and (e), respectively, (d) and (g) are the residuals between their prediction results and the mimic observed seismogram (a), respectively.}
\label{fig19}
\end{figure*}

\begin{figure*}[!t]
\centering
\includegraphics[width=1.0\textwidth]{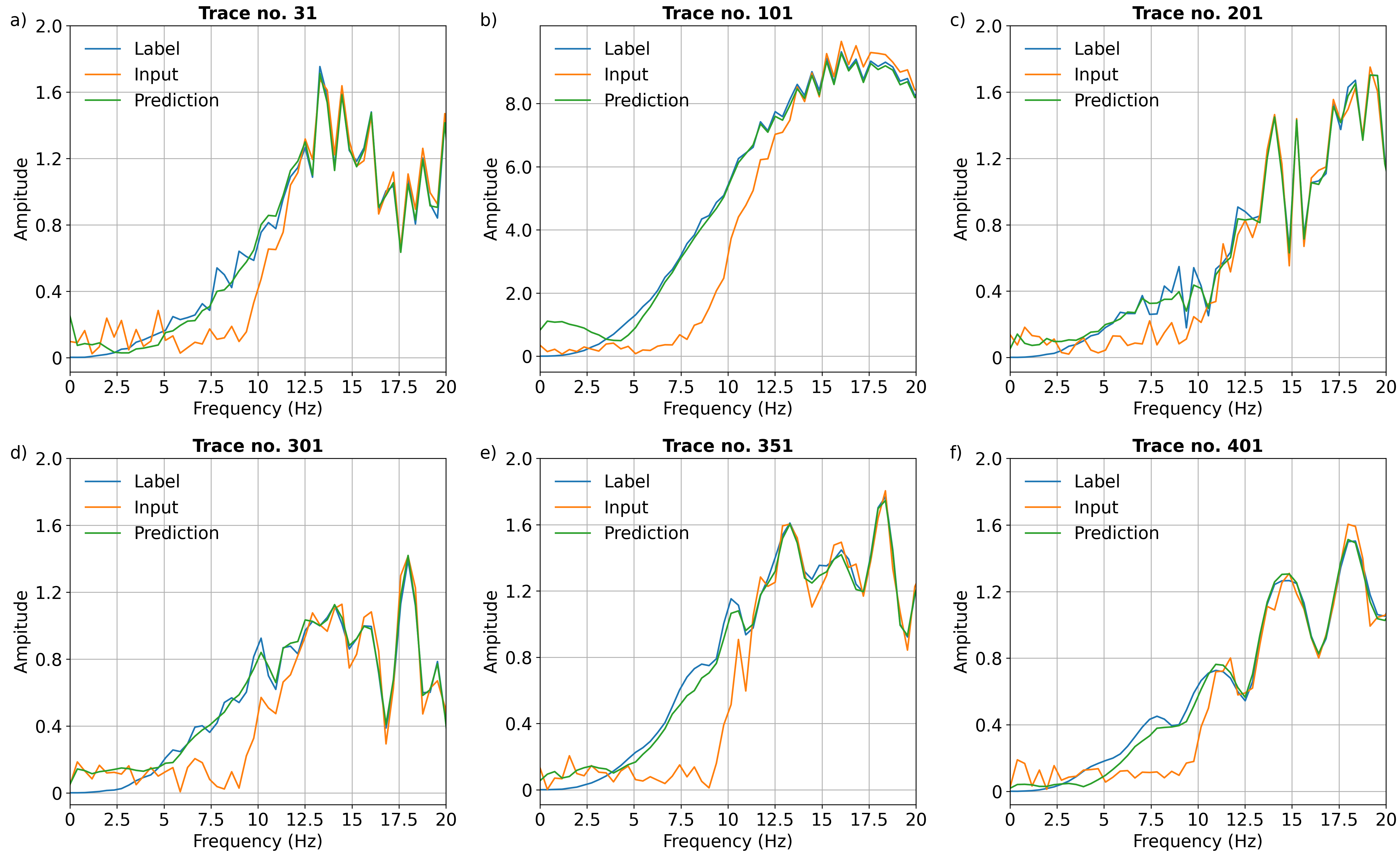}
\caption{The amplitude spectrum comparison at different locations: (a) X=0.3 km, (b) X=1 km, (c) X=2 km. (d) X=3 km, (e) X=3.5 km, and (f) X=4 km. The test input data lack frequency components below 10 Hz and contain random noise with a noise level of 30.}
\label{fig20}
\end{figure*}

\begin{figure*}[!t]
\centering
\includegraphics[width=1.0\textwidth]{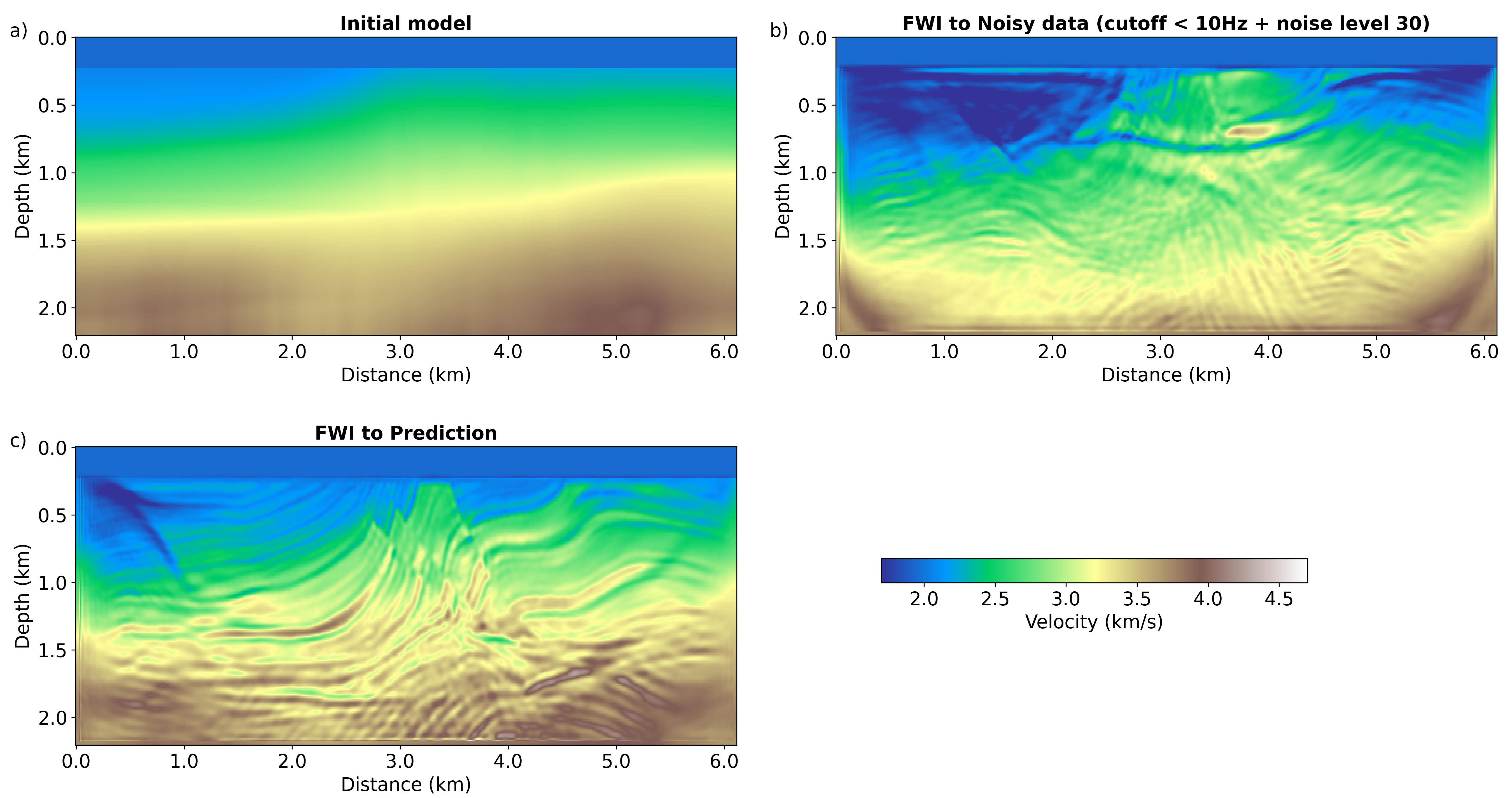}
\caption{(a) Initial velocity model. Inversion results of (b) the data with a noise level of 30 and (c) the predicted data.}
\label{fig21}
\end{figure*}

\begin{figure*}[!t]
\centering
\includegraphics[width=1.0\textwidth]{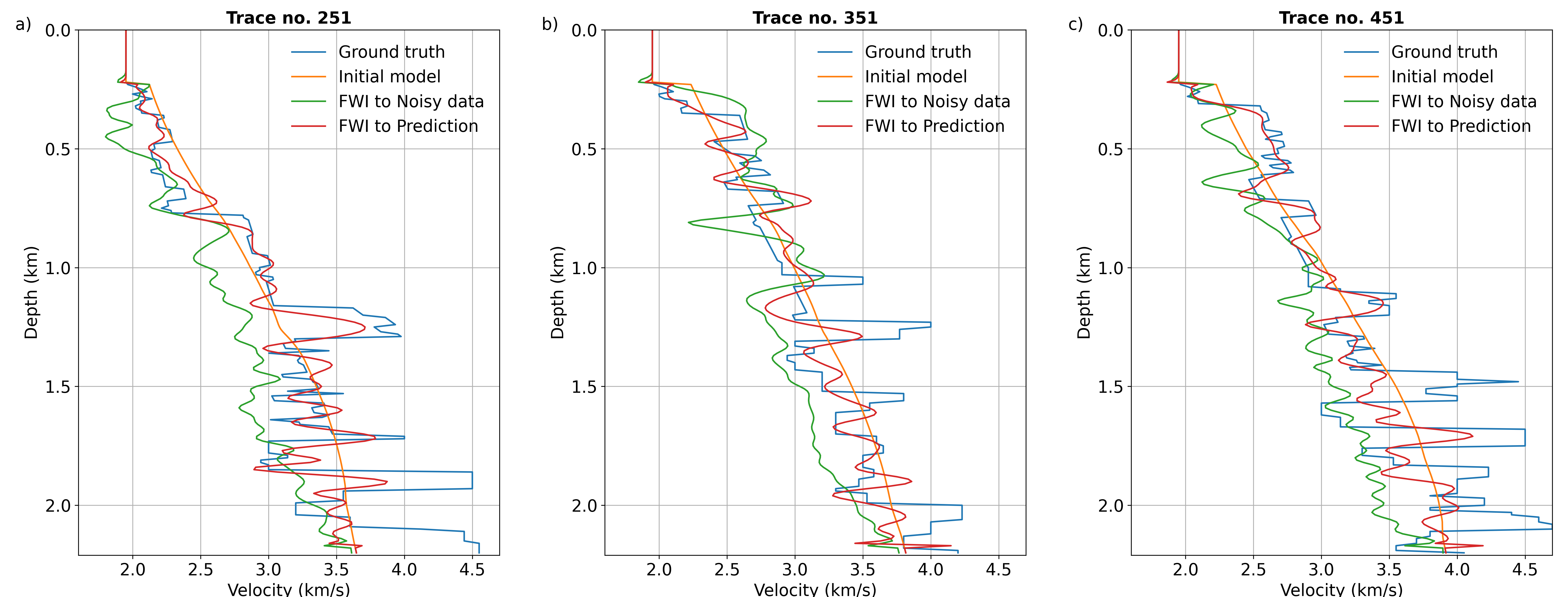}
\caption{Profiles at different locations: (a) X=2.5 km, (b) X=3.5 km and (c) X=4.50 km.}
\label{fig22}
\end{figure*}

Figure \ref{fig19} presents the test results for a noisy shot gather. Panel (a) displays the original simulated data. Panel (b) shows the data from panel (a) with frequencies below 10Hz removed and random noise of level 15 added. The corresponding prediction results are shown in panel (c), with the residuals between the prediction and the original simulated data displayed in panel (d). Compared to panel (b), panel (e) contains a shot gather with stronger noise, the corresponding prediction results for which are shown in panel (f), and the discrepancy between the prediction and the original simulated data is displayed in panel (g). It is evident that our framework is effective in removing noise. Undoubtedly, our method also attenuates some of the signal energy, particularly the deep signals in scenarios with strong noise. This occurs because we lack frequency components below 10Hz. In such instances, the added noise significantly overwhelms the weaker deep reflection signals, leading to the attenuation of these signals' energy while removing noise. To further validate the frequency extension performance, we plot the amplitude spectrum curve of the prediction (panel (f)), which is shown is Figure \ref{fig20}, comparing it with the original simulated data and the test data from panel (e). From the figures, we can see that in the frequency range below 10 Hz, the input test data exhibits random fluctuations in the amplitude spectrum due to noise contamination. However, our framework successfully removes these noise-induced artifacts and effectively extends the low-frequency components. We can clearly see that frequencies below 10 Hz have been restored. Furthermore, we make comparisons to the FWI results with different data sets, as shown in Figure \ref{fig21}. Compared to the inversion result (Figure \ref{fig21}b) of the noisy data missing the low-frequency information, the result (Figure \ref{fig21}c) of the predicted data after low-frequency extrapolation can recover the subsurface velocity favorably, especially in the middle area with adequate acquisition coverage. Figure \ref{fig22} displays the corresponding profiles at X=2.5 km, 3.5 km and 4.5 km. Obviously, the predicted inversion results (red line) match the true velocity (blue line) very well, while the noisy inversion results have a large error.

\subsection{The role of the amplitude spectrum loss}
In the process of optimizing network training, we introduce an amplitude spectrum loss to work alongside the data loss in measuring the misfit between the network output and the pseudo labels. We do this to ensure the network focuses not only on data reconstruction but also on maintaining consistency in the frequency domain, as our goal is to effectively recover low-frequency components. Integrating an amplitude spectrum loss compels the network to more accurately capture frequency characteristics during training.

Consequently, a pertinent question arises: how do we adjust the weights of the data and frequency losses? To address this issue, we conduct a test to compare how different hyperparameter $\xi$ settings influence the low-frequency extrapolation performance of our method. We employ the same training configuration as in Section 3.1, with the sole difference being the $\epsilon$ is set to 0, 0.001, 0.01, 0.1, and 1, where 0 corresponds to using only the data loss. To quantitatively assess the low-frequency extrapolation performance, we use the MAE metric to measure the discrepancy between the prediction and original simulated data. The MAE metrics for networks trained with different $\epsilon$ settings are displayed in Table \ref{tab1}. It demonstrates that incorporating the amplitude spectrum loss contributes to the enhanced performance in low-frequency extrapolation. For example, the network trained with setting $\epsilon=0.01$ achieves better performance in recovering low-frequency components than that using only data loss, as evidenced by a lower MAE metric. 

However, it's crucial to recognize that careful setting of the hyperparameter $\epsilon$ is necessary, as excessively high values can lead to a decline in network performance. For example, compared to using only data loss, the hyperparameter settings of $\epsilon=0.1$ and $\epsilon=1$ result in poorer low-frequency recovery. Based on our experience, we find that a hyperparameter setting of $\epsilon=0.01$ demonstrates the robustness across various datasets, consistently enhancing low-frequency extrapolation performance compared to using data loss alone. Therefore, we recommend setting this parameter to 0.01, as adopted in our numerical examples.

\begin{table}
\centering
\caption{The comparison of low-frequency extrapolation performance using different hyperparameter $\epsilon$, where the MAE metric has been used to measure the misfit between the prediction and the original simulated data.}
\renewcommand\arraystretch{1.5}
\setlength{\tabcolsep}{20pt}
\begin{tabular}{cccc}
    \hline
    \text {Hyperparameter setting} & \text { Cutoff 5 Hz }  & \text { Cutoff 10 Hz }  & \text { Cutoff 15 Hz } \\
    \hline
    $\epsilon = 0$ & $0.000556$ & $0.000357$ & $0.000251$ \\
    $\epsilon = 0.001$ & $0.000541$ & $0.000325$ & $0.000207$ \\
    $\epsilon = 0.01$ & $0.000486$ & $0.000286$ & $0.000198$ \\
    $\epsilon = 0.1$ & $0.000951$ & $0.000818$ & $0.000788$ \\
    $\epsilon = 1$ & $0.000619$ & $0.000473$ & $0.000217$ \\
    \hline
\end{tabular}
\label{tab1}
\end{table}

\subsection{High-pass filter setting}
Our method draws inspiration from the Noisier2Noise concept, where we create input data that lack even more low-frequency components by applying a high-pass filter to the pseudo-label data. In the process of applying this high-pass filter, we need to provide a range for high-pass cutoff frequencies. It is important to emphasize that the cutoff frequencies range is not arbitrarily set; it involves a certain trick. We will share insights on setting the cutoff frequency range, which can aid in enhancing the network's low-frequency extrapolation performance.

First, we need to analyze the frequency distribution range of the seismic data we have, as this will provide the prior knowledge required for setting the high-pass filter. Taking the synthetic data test in Section 3.1 as an example, we can determine the range of low frequencies missing in the original data to be $5\sim15$ Hz by plotting the amplitude spectrum curve. Subsequently, during the warm-up phase, our experience in setting the cutoff frequency range is guided by the principle that the maximum cutoff frequency should not exceed twice the highest missing frequency, and the minimum cutoff frequency should be slightly above the lowest missing frequency. Therefore, as observed, in Section 3.1, we set the filtering cutoff frequency range for the pseudo-labels to $6\sim30$ Hz during the warm-up phase.

In the IDR phase, the optimal approach is to gradually increase the upper limit of the filter cutoff frequency while keeping the lower limit consistent with the lowest frequency missing in the original seismic data. For example, in Section 3.1, the range of the filter cutoff frequency for the initial 50 epochs varied randomly between 5 to 7 Hz. Then, every 50 epochs, we increased the maximum filter cutoff frequency by 2 Hz. Once the maximum filter cutoff frequency aligned with the highest frequency missing in the original seismic data, we maintained this range for training until the predefined maximum training epochs. The rationale behind this is our intention to progressively enhance the network's low-frequency extrapolation performance. For example, extending the low-frequency range of data missing frequencies below 10 Hz is simpler compared to data missing frequencies below 5 Hz. Hence, we aim for the network to commence learning from these simpler tasks and then progressively increase the difficulty during training by incorporating data missing more low frequencies. This approach to learning is more stable and significantly improves the network's low-frequency extrapolation performance.

\begin{figure*}[!t]
\centering
\includegraphics[width=1.0\textwidth]{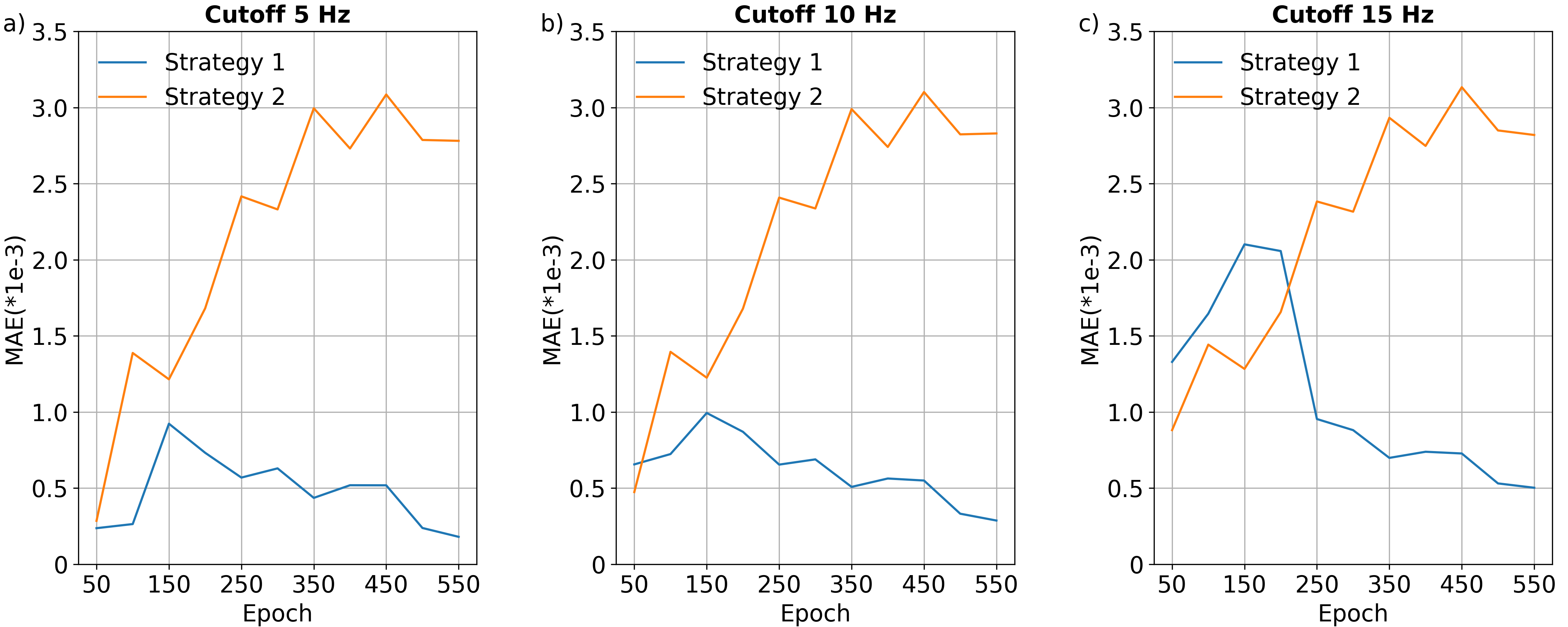}
\caption{Comparison of the low-frequency extrapolation performance of networks trained with different high-pass filter cutoff frequency setting strategies. Strategy 1 denotes gradually increasing the upper limit of the cutoff frequency during the IDR stage, while Strategy 2 involves setting a fixed cutoff frequency range. The panels (a), (b), and (c) correspond to the MAE metrics of networks trained with these two strategies on three test data, each missing frequencies below 5 Hz, 10 Hz, and 15 Hz, respectively.}
\label{fig23}
\end{figure*}

We, here, present an example to validate the effectiveness of this training strategy. Compared to Section 3.1, we only modify the settings of the filter cutoff frequency range. Specifically, in the IDR phase, we abandoned the gradually increasing cutoff frequency range setting strategy used in Section 3.1 (referred to as Strategy 1), and instead, directly set the cutoff frequency range to $5\sim15$ Hz (referred to as Strategy 2). We compare the predictions of the network models at different training epochs on three test data, each missing frequencies below 5 Hz, 10 Hz, and 15 Hz, against the network trained in Section 3.1. Figure \ref{fig23} displays the MAE metrics of the networks using two different cutoff frequency setting strategies on three test data. The panels a, b, and c correspond to the seismic data missing frequencies below 5 Hz, 10 Hz, and 15 Hz, respectively. It is evident that the network employing Strategy 1 shows gradually improving performance as the training epochs increase. In contrast, the network using Strategy 2 seems to exhibit a declining performance trend on all three test data, indicating non-convergent characteristics. These results highlight the effectiveness of our proposed training strategy, proving its significant enhancement in training stability and low-frequency extrapolation capabilities of the network.

\section{Conclusion}
We developed a novel neural network (NN)-based seismic low-frequency extrapolation method in a self-supervised learning (SSL) fashion. Under our framework, the NN sequentially undergoes two stages: warm-up and iterative data refinement (IDR). In the warm-up stage, we construct a lesslow-low dataset, using the original observed data, which lack low-frequency components, as pseudo-labels. The input data are obtained by applying a high-pass filter to these pseudo-labels, resulting in a further loss of low-frequency content. The NN rapidly warms up on this constructed dataset, initially extracting the original data's frequency characteristics and providing a degree of low-frequency extension capability. During the IDR stage, we update the lesslow-low dataset in each training epoch, where the pseudo-labels are derived from the network's predictions of the original observed data from the previous epoch, and the input data are created by applying a high-pass filter to these predicted pseudo-labels. Continually updating the training dataset allows us to gradually reduce the frequency information bias between the network's predictions and the ideal ground truth, thereby steadily enhancing the network's low-frequency extrapolation performance. We validated our method's effectiveness on both synthetic and field data in exploration circumstances. The results demonstrated that our method effectively extrapolates low-frequency components, enabling us to address the main challenge of full waveform inversion, specifically cycle-skipping. Further testing on earthquake seismogram demonstrated our method's applicability to extending ultra-low-frequency content in large-scale collected data.

\section*{Acknowledgments}
This publication is based on work supported by the King Abdullah University of Science and Technology (KAUST). The authors thank the DeepWave sponsors for supporting this research. They also thank CGG for sharing the field seismic data. This work utilized the resources of the Supercomputing Laboratory at King Abdullah University of Science and Technology (KAUST) in Thuwal, Saudi Arabia.

\bibliographystyle{unsrt}  
\bibliography{references}

\end{document}